\documentclass[useAMS,usenatbib]{mn2e}

\usepackage{amsfonts}
\usepackage{graphicx}
\usepackage{subfigure}
\usepackage{txfonts}
\usepackage{ifpdf}
\ifpdf
  \usepackage[pdftex,
    pdfauthor     = {Jason McEwen},
    pdftitle      = {Detection of the ISW effect and corresponding dark energy constraints},
    pdfsubject    = {CMB-LSS correlation},
    pdfkeywords   = {cosmic microwave background -- methods: data analysis -- methods: numerical},
    pdfproducer   = {pdflatex with hyperref},
    pdfcreator    = {pdflatex}]{hyperref}
\else
  \usepackage[ps2pdf,colorlinks=true]{hyperref}
  \hypersetup{%
    pdfauthor     = {Jason McEwen},
    pdftitle      = {Detection of the ISW effect and corresponding dark energy constraints},
    pdfsubject    = {CMB-LSS correlation},
    pdfkeywords   = {LaTeX, PDF},
    pdfauthor     = {ps2pdf with hyperref},
    pdfcreator    = {LateX},
  }
\fi

\newif\ifbw
\bwfalse

\newcommand{\eqn}[1]{(#1)}

\newcommand{\tbl}[1]{Table~#1}

\newcommand{\fig}[1]{Fig.~#1}

\newcommand{\sectn}[1]{section~#1}

\newcommand{\appn}[1]{Appendix~#1}

\newcommand{\etal}{\mbox{et al.}}
\newcommand{\eg}{\mbox{e.g.}}
\newcommand{\ie}{\mbox{i.e.}}
\newcommand{\etc}{\mbox{etc.}}

\newcommand{\astroph}{astro-ph}

\newcommand{\mnras}{Mon.\ Not.\ Roy.\ Astron.\ Soc.}
\newcommand{\physd}{\mbox{Phys.\ Rev.\ D.}}
\newcommand{\physlett}{\mbox{Phys.\ Rev.\ Lett.}}
\newcommand{\astron}{Astron. J.}
\newcommand{\apj}{ApJ}
\newcommand{\apjl}{ApJL}
\newcommand{\apjs}{ApJS}
\newcommand{\acha}{Applied and Computational Harmonic Analysis}

\newcommand{\jmp}{J.\ of Math.\ Phys.}

\newcommand{\cmb}{{CMB}}
\newcommand{\cmbtext}{{cosmic microwave background}}

\newcommand{\wmap}{{WMAP}}
\newcommand{\wmaptext}{{Wilkinson Microwave Anisotropy Probe}}

\newcommand{\cobe}{\mbox{COBE-DMR}}
\newcommand{\cobetext}{Cosmic Background Explorer-Differential Microwave Radiometer}

\newcommand{\isw}{{ISW}}
\newcommand{\iswtext}{{integrated Sachs-Wolfe}}

\newcommand{\cswt}{{CSWT}}
\newcommand{\cswttext}{continuous spherical wavelet transform}

\newcommand{\lss}{{LSS}}
\newcommand{\lsstext}{large scale structure}

\newcommand{\nvss}{{NVSS}}
\newcommand{\nvsstext}{{NRAO VLA Sky Survey}}

\newcommand{\sdss}{{SDSS}}
\newcommand{\sdsstext}{Sloan Digital Sky Survey}

\newcommand{\twomass}{{2MASS}}
\newcommand{\twomasstext}{Two Micron All Sky Survey}

\newcommand{\lcdm}{\ensuremath{\Lambda}{CDM}}

\newcommand{\smhwtext}{{spherical Mexican hat wavelet}}
\newcommand{\smhw}{{\rm SMHW}}
\newcommand{\smw}{{\rm SMW}}
\newcommand{\sbwtext}{{spherical butterfly wavelet}}
\newcommand{\sbw}{{\rm SBW}}

\newcommand{\healpix}{{HEALPix}}

\newcommand{\cmbfast}{\mbox{CMBFAST}}

\newcommand{\lambdaarch}{{LAMBDA}}
\newcommand{\lambdaarchtext}{{Legacy Archive for Microwave Background Data Analysis}}

\newcommand{\ghz}{{GHz}}

\newcommand{\nside}{\ensuremath{N_{\rm side}}}
\newcommand{\kpzero}{{Kp0}}
\newcommand{\eqdec}{\ensuremath{\delta}}

\newcommand{\olambda}{\ensuremath{\Omega_\Lambda}}
\newcommand{\w}{\ensuremath{w}}

\newcommand{\clnt}{\ensuremath{C_\el^{\rm NT, obs}}}
\newcommand{\clnttheo}{\ensuremath{C_\el^{\rm NT}}}
\newcommand{\clnntheo}{\ensuremath{C_\el^{\rm NN}}}
\newcommand{\cltttheo}{\ensuremath{C_\el^{\rm TT}}}
\newcommand{\nd}{\ensuremath{n}}
\newcommand{\tp}{\ensuremath{t}}
\newcommand{\ndlab}{\ensuremath{{\rm N}}}
\newcommand{\tplab}{\ensuremath{{\rm T}}}

\newcommand{\degrees}{\ensuremath{{^\circ}}}

\newcommand{\snr}{\ensuremath{{\rm SNR}}}

\newcommand{\wcov}{\ensuremath{X_\wav}}
\newcommand{\wcovest}{\ensuremath{\hat{X}_\wav}}

\newcommand{\wcovvect}{\ensuremath{\mathbf{X}_\wav}}
\newcommand{\wcovestvect}{\ensuremath{\hat{\mathbf{X}}_\wav}}

\newcommand{\weight}{\ensuremath{\nu}}

\newcommand{\covsubterm}{\ensuremath{G^\el}}

\newcommand{\chisqd}{\ensuremath{\chi^2}}
\newcommand{\baypar}{\ensuremath{\Theta}}
\newcommand{\lhood}{\ensuremath{\mathcal{L}}}

\newcommand{\dndz}{\ensuremath{\frac{\dx N}{\dx z}}}
\newcommand{\dgdz}{\ensuremath{\frac{\dx g}{\dx z}}}

\newcommand{\kron}{\ensuremath{\delta}}

\newcommand{\dotprod}{\ensuremath{\cdot}}

\newcommand{\pixw}{\ensuremath{p_\el}}
\newcommand{\bm}[1]{\ensuremath{b_\el^{#1}}}

\newcommand{\el}{\ensuremath{\ell}}
\newcommand{\m}{\ensuremath{m}}
\newcommand{\elm}{\ensuremath{{\el\m}}}

\newcommand{\elmax}{\ensuremath{{\el_{\rm max}}}}

\newcommand{\shfarg}[3]{\ensuremath{Y_{#1#2}({#3})}}
\newcommand{\shfargc}[3]{\ensuremath{Y_{#1#2}^\cconj({#3})}}

\newcommand{\leg}[2]{\ensuremath{P_{{#1}}({#2})}}

\newcommand{\sbessel}{\ensuremath{j}}

\newcommand{\xvect}{\ensuremath{\mathbf{x}}}
\newcommand{\kvect}{\ensuremath{\mathbf{k}}}
\newcommand{\knaut}{\ensuremath{k_0}}

\newcommand{\spcend}{\ensuremath{\:}}

\newcommand{\img}{\ensuremath{\mathit{i}}}

\newcommand{\dx}{\ensuremath{\mathrm{\,d}}}
\newcommand{\dmu}[1]{\ensuremath{\dx \Omega(#1)}}

\newcommand{\cconj}{\ensuremath{\ast}}  

\newcommand{\sky}{\ensuremath{s}}
\newcommand{\skywav}{\ensuremath{{W}}}

\newcommand{\sa}{\ensuremath{\omega}}
\newcommand{\saa}{\ensuremath{\theta}}
\newcommand{\sab}{\ensuremath{\phi}}
\newcommand{\sas}{\ensuremath{\saa, \sab}}
\newcommand{\eul}{\ensuremath{\mathbf{\rho}}}
\newcommand{\euls}{\ensuremath{\eula, \eulb, \eulc}}
\newcommand{\eula}{\ensuremath{\alpha}}
\newcommand{\eulb}{\ensuremath{\beta}}
\newcommand{\eulc}{\ensuremath{\gamma}}

\newcommand{\spo}{\ensuremath{\Pi}}
\newcommand{\cocycle}{\ensuremath{\lambda}}

\newcommand{\realno}{\ensuremath{\mathbb{R}}}

\newcommand{\sothree}{\ensuremath{\mathrm{SO}(3)}}
\newcommand{\sphere}{\ensuremath{{S^2}}}
\newcommand{\dil}{\ensuremath{\mathcal{D}}}
\newcommand{\dilsmall}{\ensuremath{d}}
\newcommand{\rot}{\ensuremath{R}}
\newcommand{\scalea}{\ensuremath{a}}
\newcommand{\scaleb}{\ensuremath{b}}
\newcommand{\scaleab}{\ensuremath{{\scalea,\scaleb}}}

\newcommand{\effsize}{\ensuremath{\xi}}

\newcommand{\p}{\ensuremath{^\prime}}
\newcommand{\pp}{\ensuremath{^{\prime\prime}}}
\newcommand{\wav}{\ensuremath{\psi}}

\newcommand{\nsigma}{\ensuremath{N_\sigma}}

\newcommand{\opnexpv}{\ensuremath{\left<}}
\newcommand{\clsexpv}{\ensuremath{\right>}}

\title[Detection of the ISW effect and corresponding dark energy constraints]
  {Detection of the ISW effect and corresponding dark energy constraints %
   made with directional spherical wavelets}

\author[J.~D.~McEwen \etal]
  {J.~D.~McEwen$^1$\thanks{E-mail: mcewen@mrao.cam.ac.uk}, P.~Vielva$^{2,3}$,
    M.~P.~Hobson$^1$, E.~Mart\'{\i}nez-Gonz\'{a}lez$^2$, 
    A.~N.~Lasenby$^1$\\
  $^1$Astrophysics Group, 
      Cavendish Laboratory,  J.~J.~Thomson Avenue,
      Cambridge CB3 0HE, UK\\
  $^2$Instituto de F\'{\i}sica de Cantabria,
      {CSIC-Universidad de Cantabria}, Avda.\ los Castros s/n,
      39005, Santander, Spain\\
  $^3$Laboratoire APC, Coll\`{e}ge de France,
      F-75231, Paris Cedex 5, France}
\date{Accepted ---. Received ---; in original form \today}
\pagerange{\pageref{firstpage}--\pageref{lastpage}} 
\pubyear{2005}

\def\LaTeX{L\kern-.36em\raise.3ex\hbox{a}\kern-.15em
    T\kern-.1667em\lower.7ex\hbox{E}\kern-.125emX}

\begin{document}
\label{firstpage}
\maketitle


\begin{abstract}
Using a directional spherical wavelet analysis we detect the \iswtext\
(\isw) effect, indicated by a positive correlation between the first-year
\wmaptext\ (\wmap) and \nvsstext\ (\nvss) data.
Detections are made using both a directional extension
of the
\smhwtext\ and the \sbwtext.
We examine the
possibility of foreground contamination and systematics in the \wmap\
data and conclude that these factors are not responsible for the
signal that we detect.  The wavelet analysis inherently enables us to
localise on the sky those regions that contribute most strongly to the
correlation.  On removing these localised regions the correlation that
we detect is reduced in significance, as expected, but it is not
eliminated, suggesting that these regions are not the sole source of
correlation between the data.  This finding is consistent with
predictions made using the \isw\ effect, where one would expect weak
correlations over the entire sky.  
%
%
In a flat universe the detection of the \isw\ effect provides direct
and independent evidence for dark energy.  We use our detection
to constrain dark energy parameters by deriving a theoretical
prediction for the directional wavelet covariance statistic for a
given cosmological model.  Comparing these predictions with the data
we place constraints on the equation-of-state parameter \w\ 
and the vacuum energy
density \olambda.  
We also consider the case of a pure cosmological constant, \ie\
$\w=-1$.  
For this case we 
rule out a zero cosmological
constant at greater than the 99.9\% significance level.
%
All parameter estimates that we obtain are consistent with the standand cosmological concordance model values.
Although wavelets perform very well when attempting to detect the ISW
effect since one may probe only the regions where the signal is
present, once all information is incorporated when computing parameter
estimates, the performance of the wavelet analysis is comparable to
other methods, as expected for a linear approach.
\end{abstract}




\section{Introduction}

Strong observational evidence now exists in support of the $\Lambda$ cold dark matter (\lcdm) fiducial model of the universe; specifically, a universe that is (nearly) flat and dominated by an exotic dark energy component.  We know very little about the origin and nature of this dark energy, but we now have strong evidence in support of its existence and relative abundance.
Much of this evidence comes from recent measurements of the \cmbtext\ (\cmb) anisotropies of high resolution and precision, in particular the recent \wmaptext\ (\wmap) data \citep{bennett:2003a}.  The existence of dark energy has also been independently found by measurements of the luminosity distance to Supernova Type Ia
\citep{riess:1998,perlmutter:1999}.

At this point, the confirmation of the fiducial \lcdm\ model and the existence of dark energy by independent physical methods is of particular interest.
One such approach is through the detection of the \iswtext\ (\isw) effect \citep{sachs:1967}.
\cmb\ photons are blue and red shifted as they fall into and out of gravitational potential wells, respectively, as they travel towards us from the surface of last scattering.  If the gravitational potential evolves during the photon propagation, then the blue and red shifts do not cancel exactly and a net change in the photon energy occurs.  This secondary induced \cmb\ anisotropy (the \isw\ effect) exists only in the presence of spatial curvature or, in a flat universe, in the presence of dark energy \citep{peebles:2003}. The recent \wmap\ 1-year data has imposed strong constraints on the flatness of the universe \citep{bennett:2003a,spergel:2003}, hence any \isw\ signal may be interpreted directly as a signature of dark energy.

It is difficult to separate directly the contribution of the \isw\ effect from the \cmb\ anisotropies, hence it is not feasible to detect the \isw\ effect solely from the \cmb.  Instead, as first proposed by \citet{crittenden:1996}, the \isw\ effect may be detected by cross-correlating the \cmb\ anisotropies with tracers of the local matter distribution, such as the nearby galaxy density distribution.
A detection of large-scale positive correlations is a direct indication of the \isw\ effect, and correspondingly, evidence for dark energy.

The first attempt to detect the cross-correlation between the \cmb\ and the nearby galaxy distribution was performed by \citet{boughn:2002} using the \cobetext\ (\cobe; \citealt{bennett:1996}) and \nvsstext\ (\nvss; \citealt{condon:1998}) data.%
\footnote{As the detection of the large-scale effect is cosmic variance limited, one requires (near) full-sky maps.}
No cross-correlation was found; \citet{boughn:2002} conclude that a future experiment with better sensitivity and resolution was required to make any detection.
The \wmap\ mission has now provided a suitable experiment, and several groups have since reported detections of the \isw\ effect at a range of significance levels using various tracers of the local universe.
\citet{boughn:2004,boughn:2005} cross-correlate the \wmap\ data with two different tracers of the nearby universe: the hard X-ray data provided by the
High Energy Astronomy Observatory-1 satellite (HEAO-1; \citealt{boldt:1987}) and the \nvss\ data.  They make a statistically significant detection of the cross-correlation at the $1.8$--$2.8\sigma$ level at scales below $3^\circ$.
\citet{nolta:2004} perform an independent analysis of the \wmap\ and \nvss\ data and confirm the existence of dark energy at the $95$\% significance level.  
\citet{fosalba:2004} cross-correlate the \wmap\ data with the {APM} galaxy survey \citep{maddox:1990} and report a cross-correlation detection at the $2.5\sigma$  level
on scales of $4^\circ$--$10^\circ$.
\citet{fosalba:2003} cross-correlate the \wmap\ data with the \sdsstext\ (\sdss; \citealt{york:2000}) and detect an \isw\ signal at the $3\sigma$ level.
These same two data sets are also analysed by \cite{scranton:2003}.\footnote{See \citet{afshordietal:2004} for a critical discussion of the analyses done by \citet{fosalba:2004}, \citet{fosalba:2003} and \citet{scranton:2003}.}
\citet{afshordietal:2004} cross-correlate the \wmap\ data with the near infrared \twomasstext\ Extended Source Catalogue (\twomass\ {XSC}; \citealt{jarrett:2000}) and detect an \isw\ signal at the $2.5\sigma$ level.
\citet{padmanabhan:2004} make a $2.5\sigma$ detection of the \isw\ effect by cross-correlating the \wmap\ and \sdss\ data, and subsequently use the detection to constrain cosmological parameters.  Other works have focused on the theoretical detectability of the \isw\ effect for various experiments and, in some cases, the use of such detections to constrain cosmological parameters \citep{afshordi:2004,hu:2004,pogosian:2004,pogosian:2005,corasaniti:2005}.

The previous works discussed all perform the cross-correlation of the \cmb\ with various tracers of the near universe \lsstext\ (\lss) in either real or harmonic space, using the real space angular correlation function or the cross-angular power spectrum respectively.  Recently, \citet{vielva:2005} adopt a different measure by performing the cross-correlation in spherical wavelet space using the azimuthally symmetric \smhwtext\ (\smhw).
Spherical wavelets have already been used in many astrophysical and cosmological applications.  For example, spherical wavelets have been used extensively to test the \cmb\ for non-Gaussianity and isotropy \citep{barreiro:2000,cayon:2001,cayon:2003,cayon:2005,martinez:2002,vielva:2004,mukherjee:2004,cruz:2005,cruz:2006,mcewen:2005b,liu:2005,wiaux:2006}.
However, \citet{vielva:2005} present the first use of spherical wavelets for cross-correlating the \cmb\ with tracers of \lss\ in an attempt to detect the \isw\ effect.
Since the \isw\ effect is localised to certain (large) scales on the sky,
wavelets are an ideal tool for searching for cross-correlations due to the inherent scale and spatial localisation afforded by a wavelet analysis.
\citet{vielva:2005} examine the covariance of the \smhw\ coefficients of the \wmap\ and \nvss\ data for a range of scales, making a detection of the \isw\ effect at the $3.3\sigma$ level at scales on the sky of $6^\circ$--$8^\circ$.  Moreover, the detection is used to constrain cosmological parameters that describe the dark energy.

There is no physical reason to assume that the local correlated structures induced in the \cmb\ anisotropies by the near \lss\ are rotationally invariant; indeed, it is known that Gaussian random fields are characterised by features that are not necessarily rotationally invariant \citet{barreiro:1997,barreiro:2001}.  Thus, other spherical wavelets that are not azimuthally symmetric, \ie\ directional\footnote{We use the term `directional wavelet' to refer to a wavelet that is not azimuthally symmetric.  This may differ to the definition of a directional wavelet used by some other authors.} wavelets, may be equally (or more) suitable for probing the data for cross-correlations.  Herein we extend the analysis performed by \citet{vielva:2005}, using directional wavelets to examine the cross-correlation of the \wmap\ and \nvss\ data.  We use any detections of the \isw\ made to constrain cosmological parameters that describe the dark energy.

The remainder of this paper is structured as follows.  The directional \cswttext\ and the wavelets that we consider are described in \sectn{\ref{sec:cswt}}.  In \sectn{\ref{sec:wcov}} we define the wavelet covariance estimator used to test for correlations and also define the theoretical covariance predicted for a given cosmological model (the derivation for which is presented in \appn{\ref{appn:theo_covar}}).  
We then 
compare the expected performance of various wavelets for detecting the \isw\ effect.  In \sectn{\ref{sec:analysis}} we give a brief overview of the data considered and the analysis procedure.  Results are presented and discussed in \sectn{\ref{sec:results}}.  We describe the detections made and then use the detections to place constraints on dark energy parameters.  Concluding remarks are made in \sectn{\ref{sec:conclusions}}


\section{Continuous spherical wavelet transform}
\label{sec:cswt}

To perform a wavelet analysis of full-sky maps defined on the celestial sphere, Euclidean wavelet analysis must be extended to spherical geometry.
A wavelet transform on the sphere has already been constructed and developed  \citep{antoine:1998,antoine:1999,antoine:2002,antoine:2004}.\footnote{This framework has been extended to wavelet frames on the sphere also by \citet{bogdanova:2004}.}
The spherical wavelet construction is derived entirely from group theoretic principles, however recently \citet{wiaux:2005} reintroduce the formalism in an equivalent, practical and self-consistent approach that is independent of the original group theoretic framework.  We adopt this latter approach and 
extend the decomposition to anisotropic dilations \citep{mcewen:2005a}.  Moreover, we also apply a fast algorithm for performing the wavelet transform on the sphere \citep{mcewen:2005a}, based on the fast spherical convolution algorithm developed by \citet{wandelt:2001}.  We present here a brief overview of the \cswttext\ (\cswt), but refer the reader to our recent work \citep{mcewen:2005a} for more details on the analysis and fast algorithms.

\subsection{Wavelet transform}

The correspondence principle between spherical and Euclidean wavelets developed by \cite{wiaux:2005}, relates the concepts of planar Euclidean wavelets to spherical wavelets through a stereographic projection.  The stereographic projection is used to define affine transformations on the sphere that facilitate the construction of a wavelet basis on the sphere.  The spherical wavelet transform may then be defined as the projection on to this basis, where the spherical wavelets must satisfy the appropriate admissibility criterion to ensure perfect reconstruction.

\begin{figure}
\centerline{
  \includegraphics[width=75mm]{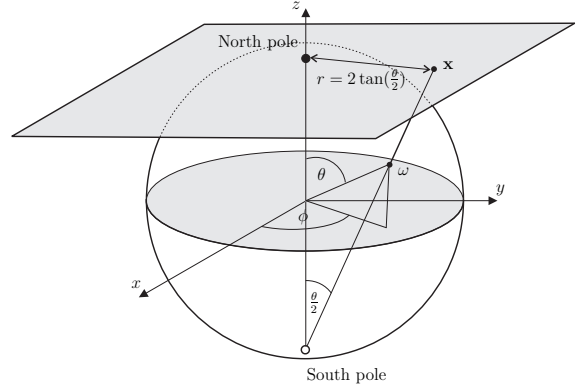}}
\caption[Stereographic projection]{Stereographic projection of the
  sphere onto the plane.}
\label{fig:stereographic_projection}
\end{figure}

The stereographic projection is defined by projecting a point on the unit sphere to a point on the tangent plane at the north pole, by casting a ray though the point and the south pole.  The point on the sphere is mapped on to the intersection of this ray and the tangent plane (see
\fig{\ref{fig:stereographic_projection}}).  The stereographic projection is radial and conformal \citep{wiaux:2005}, hence local angles are preserved under the transform, \ie\ directionality is preserved.
The stereographic projection operator that preserves the $L^2$-norm of functions is denoted $\spo$, with inverse $\spo^{-1}$.


The stereographic projection operator may be used to extend the concept of dilations to the sphere.  In particular, we adopt here the extension of the isotropic dilation defined by \citet{wiaux:2005} to anisotropic dilations \citep{mcewen:2005a}.\footnote{A similar anisotropic dilation operator on the sphere has also been independently proposed by \citet{tosic:2005}.}
Dilations on the sphere are constructed by first projecting the sphere on to the plane using the stereographic projection, performing the usual Euclidean dilation in the plane, before re-projecting back on to the sphere using the inverse stereographic projections.  The spherical dilation is thus defined by
\begin{equation}
\dil(\scaleab) = \spo^{-1} \, \dilsmall(\scaleab) \, \spo
\spcend ,
\end{equation}
where $\dilsmall(\scaleab)$ is the anisotropic Euclidean dilation on the plane, defined for a function on the plane $p \in L^2(\realno^2)$ by
$[\dilsmall(\scalea,\scaleb) p] (x,y) = \scalea^{-1/2}\scaleb^{-1/2}
p(\scalea^{-1}x,\scaleb^{-1}y)$,
for the non-zero positive scales 
$\scalea, \scaleb \in \realno^{+}_{\ast}$.
The $L^2$-norm of functions is preserved by the spherical dilation operator since both the stereographic projection and the Euclidean dilation preserve the norm.
We therefore obtain the following definition of the spherical dilation of a square-integrable function on the sphere $s \in L^2(\sphere)$:
\begin{equation}
[\dil(\scaleab) s](\sa)
= [\cocycle(\scaleab,\sas)]^{1/2} \: s(\sa_{1/\scalea,1/\scaleb})
\spcend ,
\end{equation}
where $\sa_\scaleab=(\saa_\scaleab,\sab_\scaleab)$,
\begin{displaymath}
\tan(\saa_\scaleab/2) = 
\tan(\saa/2)
\sqrt{\scalea^2 \cos^2{\sab} + \scaleb^2 \sin^2{\sab}}
\end{displaymath}
and 
$\tan(\sab_\scaleab) = 
\frac{\scaleb}{\scalea}
\tan(\sab)$.
The spherical coordinates with colatitude $\saa$ and longitude $\sab$ are denoted by $\sa=(\sas)\in\sphere$.
The $\cocycle(\scaleab,\sas)$ cocycle term follows from the various factors
introduced to preserve the $L^2$-norm of functions.
The cocycle of an anisotropic spherical dilation is defined by
\begin{equation}
\cocycle(\scaleab, \sas) =
\frac{4 \scalea^3 \scaleb^3}
{ 
\left(
A_{-}\cos\saa + A_{+}
\right)^2
}
\spcend ,
\end{equation}
where 
\begin{displaymath}
A_\pm = \scalea^2\scaleb^2 \pm \scalea^2 \sin^2\sab \pm \scaleb^2 \cos^2\sab
\spcend .
\end{displaymath}

Finally, we require an extension of Euclidean translations to the sphere in order to construct a wavelet basis on the sphere.
The natural extension of translations to the sphere are rotations.  These are characterised by the elements of the rotation group \sothree, which we parameterise in terms of the three Euler angles
$\rho=(\euls)$.\footnote{We adopt the $zyz$ Euler convention corresponding to the rotation of a physical body in a \emph{fixed} co-ordinate system about the $z$, $y$ and $z$ axes by \eulc, \eulb\ and \eula\ respectively.}  The rotation of a function on the sphere is defined by
\begin{equation}
[\rot(\rho) s](\sa) = s(\rho^{-1} \sa), \; \; \rho \in \sothree 
\spcend .
\end{equation}

A wavelet basis on the sphere may now be constructed from rotations and
dilations of a mother spherical wavelet $\wav \in L^2(\sphere)$.  
The corresponding wavelet \mbox{family} on the sphere
\mbox{$\{ \wav_{\scaleab,\rho} \equiv \rot(\eul) \dil(\scaleab) \wav;
\; {\rho \in \sothree,} \; {\scaleab \in \realno_{\ast}^{+}} \}$}
provides an over-complete set of functions in $L^2(\sphere)$.
The \cswt\ of a function on the sphere is given by the projection on to each wavelet basis function in the usual manner,
\begin{equation}
\skywav_\wav(\scaleab, \eul) 
\equiv
\int_{\sphere}
\dmu{\sa} \:
\wav_{\scaleab,\eul}^\cconj(\sa) \:
\sky(\sa)
\spcend ,
\label{eqn:cswt}
\end{equation}
where the \cconj\ denotes complex conjugation and $\dmu{\sa} = \sin\saa \dx\saa \dx\sab$ is the usual rotation invariant measure on the sphere.
In this work we do not consider the synthesis of a function on the sphere from its wavelet coefficients.  Indeed, it is only possible to synthesise a function from its spherical wavelet coefficients for the case of isotropic dilations \citep{mcewen:2005a}.

The transform is general in the sense that all orientations in the
rotation group \sothree\ are considered, thus directional structure is
naturally incorporated.  It is important to note, however, that only
\emph{local} directions make any sense on \sphere.  There is no global
way of defining directions on the sphere -- there will always be some singular point where the definition fails.

\subsection{Mother spherical wavelets}
\label{sec:mother_wavelets}

Admissible mother spherical wavelets may be constructed from the stereographic projection of admissible mother Euclidean wavelets on the plane:
\begin{equation}
\label{eqn:wav_proj}
\wav(\sa) = [\spo^{-1}\wav_{\realno^2}](\sa)
\spcend .
\end{equation}
Directional spherical wavelets may be naturally constructed in this 
setting -- they are simply the projection of directional Euclidean planar
wavelets on to the sphere.

We consider three spherical wavelets in our subsequent cross-correlation analysis: the spherical Mexican hat wavelet (\smhw); the spherical butterfly wavelet (\sbw); and the spherical real Morlet wavelet (\smw).  These spherical wavelets are illustrated in \fig{\ref{fig:mother_wavelets}}.
Each spherical wavelet is constructed by the stereographic projection
of the corresponding Euclidean wavelet onto the sphere, where the
Euclidean planar wavelets are defined by 
\begin{displaymath}
\wav_{\realno^2}^{\rm \smhw}(r,\sab) = 
\frac{1}{2}
(2 - r^2) \, {\rm e}^{-r^2/2}
\spcend ,
\end{displaymath}
\begin{displaymath}
\wav_{\realno^2}^{\rm \sbw}(x,y) = 
x \, {\rm e}^{-(x^2+y^2)/2}
\end{displaymath}
and
\begin{displaymath}
\wav_{\realno^2}^{\rm \smw}(\xvect; \kvect) = 
{\rm Re} \left(
{\rm e}^{ \img \kvect \cdot \xvect / \sqrt{2}} \:
{\rm e}^{ - \| \xvect \|^2 /2}
\right)
\end{displaymath}
respectively, where $\kvect$ is the wave vector of the \smw. 
The \smhw\ is proportional to the Laplacian of a Gaussian,
whereas the \sbw\ is proportional to the first partial derivative of a
Gaussian in the $x$-direction.  The \smw\ is a Gaussian modulated
sinusoid, or Gabor wavelet.

The dilation parameter of each wavelet may be related to an effective size on the size for the wavelet.
For directional wavelets we define two effective sizes on the sky defined in orthogonal directions.
The \smhw\ and \sbw\ are both derived from a parent Gaussian function and thus for each direction have the same effective size on the sky, defined by
$\effsize^{\smhw, \,\sbw}_i(c_i) \approx 2 \sqrt{2} \: c_i$,
where $i=\{1,2\}$ and $c_i=\{\scalea,\scaleb\}_{i=1,2}$ \citep{mcewen:2005b}.
The first effective size of the \smw\ defines the overall size of the wavelet on the sky
$\effsize^{\smw}_1(\scalea) \approx 2 \sqrt{2} \: \scalea$,
whereas the second orthogonal size defines the size of the internal structure of the wavelet
$\effsize^{\smw}_2(\scaleb) \approx \scaleb \pi / \knaut$ \citep{mcewen:2005b}.


\newlength{\mapwidth}
\setlength{\mapwidth}{60mm}

\begin{figure}
\centering
\ifbw
  \subfigure[Spherical Mexican hat wavelet (\smhw)]{\includegraphics[width=\mapwidth]{figures_bw/mexh_d0-2_moll_bw}}
  \subfigure[Spherical butterfly wavelet (\sbw)]{\includegraphics[width=\mapwidth]{figures_bw/bfly_d0-2_moll_bw}}
  \subfigure[Spherical real Morlet wavelet (\smw)]{\includegraphics[width=\mapwidth]{figures_bw/mrlt_d0-2_moll_bw}}
\else
  \subfigure[Spherical Mexican hat wavelet (\smhw)]{\includegraphics[width=\mapwidth]{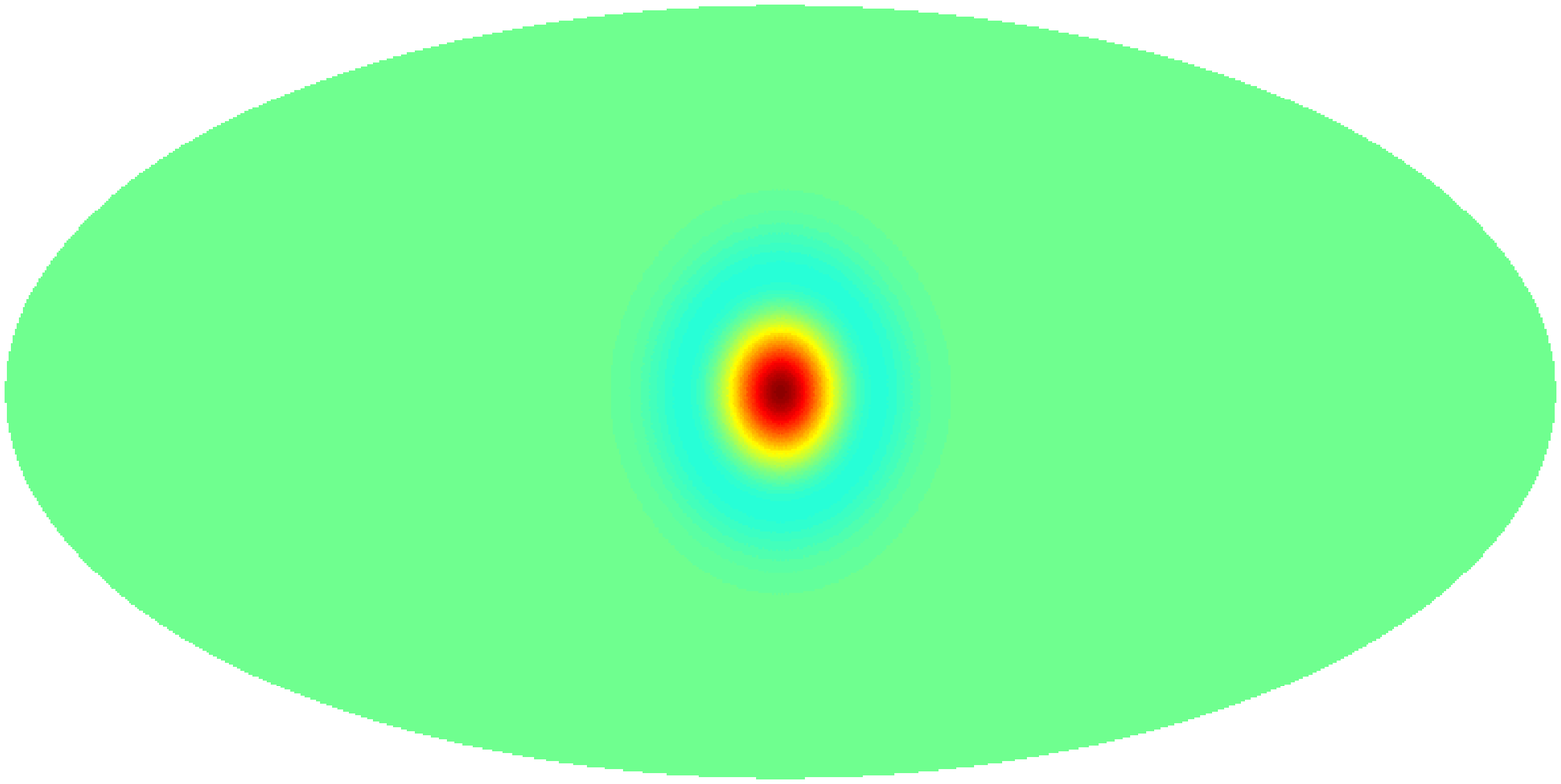}}
  \subfigure[Spherical butterfly wavelet (\sbw)]{\includegraphics[width=\mapwidth]{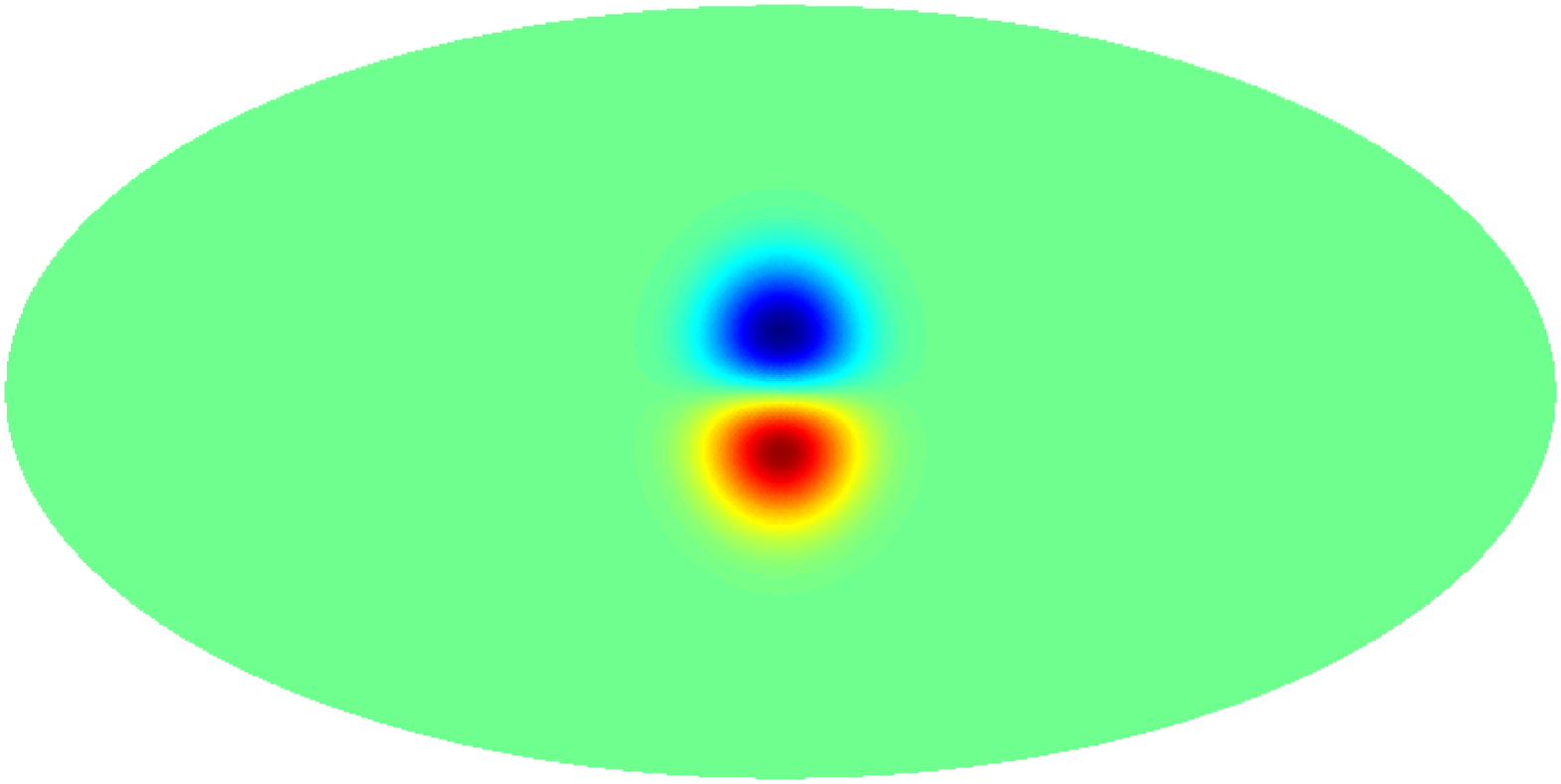}}
  \subfigure[Spherical real Morlet wavelet (\smw)]{\includegraphics[width=\mapwidth]{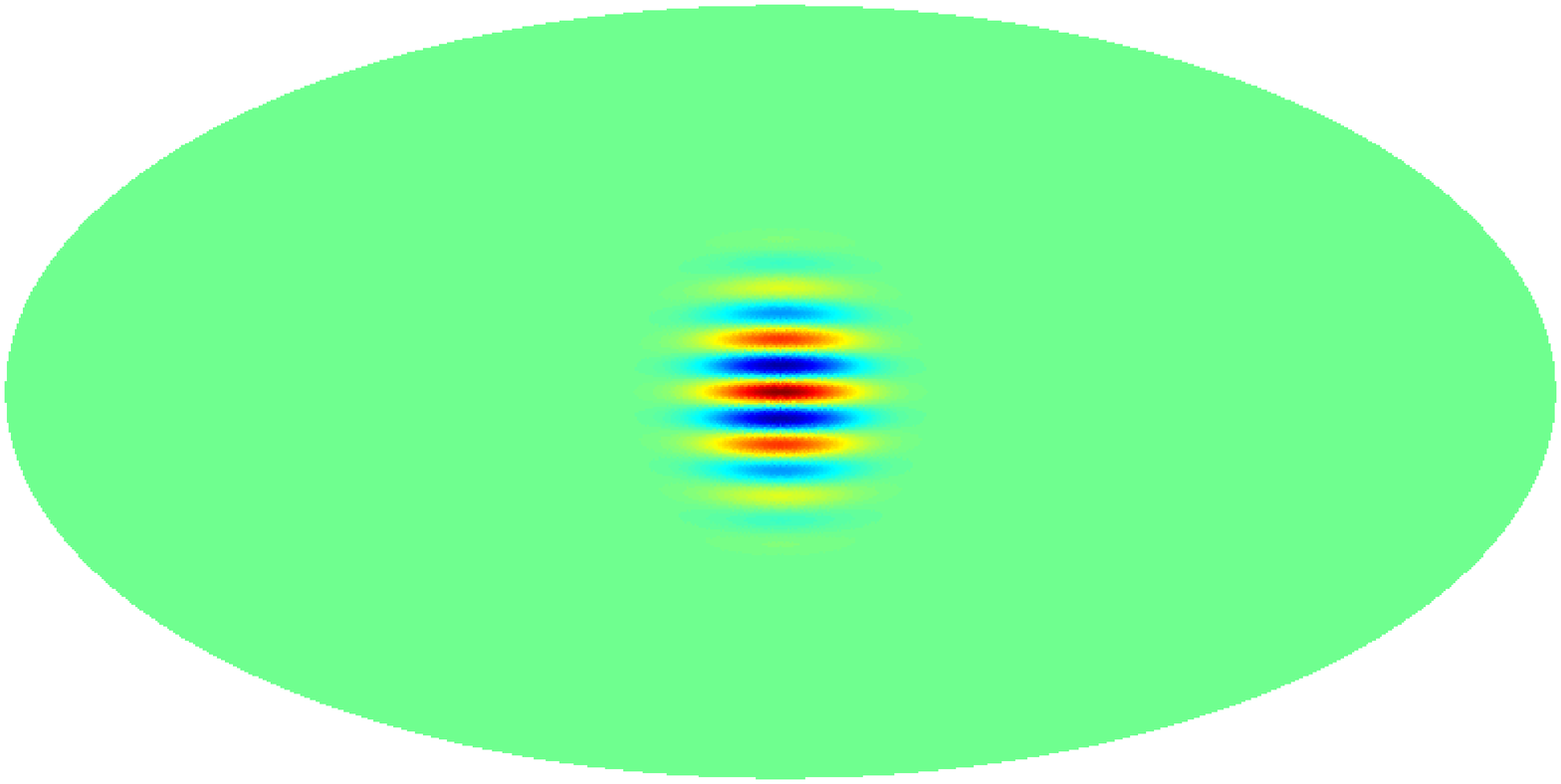}}
\fi 
\caption{Spherical wavelets at scale $\scalea=\scaleb=0.2$.  Wavelet
  maps are displayed in the Mollweide projection, where the wavelets
  have been rotated down from the north pole for ease of observation.
  The \smw\ is plotted for wave vector $\kvect=(10,0)^T$.}
\label{fig:mother_wavelets}
\end{figure}

\section{Cross-correlation in wavelet space}
\label{sec:wcov}

The effectiveness of using a spherical wavelet estimator to detect cross-correlations between the \cmb\ and the nearby galaxy density distribution has been demonstrated by \citet{vielva:2005}.  We extend the wavelet estimator here to account for directional wavelets that are not azimuthally symmetric.  Firstly, we review the theoretical cross-power spectrum of the \cmb\ and galaxy density and give an expression for the spectrum based on the particular cosmological model.  We then define the wavelet coefficient covariance estimator used to test for cross-correlations.
Using the theoretical cross-power spectrum previously described, we give a theoretical prediction for the wavelet coefficient covariance for a particular cosmological model (\ie\ for a particular theoretical cross-power spectrum).
In subsequent sections, we use this result to compare theoretical predictions of the wavelet covariance for a range of cosmologies with measurements from the data, in order to place constraints on the cosmological parameters that define the dark energy.
Finally, we also examine in this section the effectiveness of various spherical wavelets for detecting cross-correlations.

\subsection{Theoretical cross-power spectrum}
\label{sec:theo_crossspec}

For a particular cosmology, we consider the theoretical cross-power spectrum \clnttheo\ of the galaxy density map $\nd(\sa)$ with the \cmb\ temperature anisotropy map $\tp(\sa)$, defined as the ensemble average of the product of the spherical harmonic coefficients of the two maps:
\begin{equation}
\label{eqn:cltheo}
\opnexpv \nd_{\el\m}^{} \: \tp_{\el\p \m\p}^\cconj \clsexpv =
\kron_{\el\el\p} \kron_{\m\m\p} \:
\clnttheo
\spcend ,
\end{equation}
where $\kron_{ij}$  is the Kronecker delta function.
In defining the cross-correlation in this manner we implicitly assume that the galaxy density and \cmb\ random fields on the sphere are homogeneous and isotropic.
Modelling the observation process, the observed cross-spectrum for the given cosmology is related to the theoretical one by
\begin{equation}
\label{eqn:clobs}
\clnt = \pixw{}^2 \: \bm{\ndlab} \: \bm{\tplab} \: \clnttheo
\spcend ,
\end{equation}
where $\pixw$ is the pixel window function for the pixelisation scheme adopted and $\bm{\ndlab}$ and $\bm{\tplab}$ are the beam window functions for the galaxy density\footnote{There is no beaming for the \nvss\ galaxy density map subsequently used, hence $\bm{\ndlab}=1$,  $\forall \el$.} and \cmb\ maps respectively.  Pixel window and beam functions are represented here by Legendre coefficients (rather than spherical harmonic coefficients).

The theoretical cross-power spectrum may be computed for a given cosmology by (\eg\ \citealt{nolta:2004})
\begin{equation}
\clnttheo = 12 \pi \Omega_{\rm m} H_0{}^2
\int \frac{\dx k}{k^3} \:
\Delta_\delta^2(k) \:
F_\el^{\rm N}(k) \:
F_\el^{\rm T}(k)
\spcend ,
\end{equation}
where $\Omega_{\rm m}$ is the matter density, $H_0$ is the Hubble parameter, \mbox{$\Delta_\delta^2(k) = k^3P_\delta(k)/2\pi^2$} is the logarithmic matter power spectrum ($P_\delta(k)$ is the matter power spectrum) and $F_\el^{\ndlab}(k)$ and $F_\el^{\tplab}(k)$ are the filter functions for the galaxy density distribution and \cmb\ respectively, given by
\begin{equation}
F_\el^{\ndlab}(k) = b \int \dx z \: \dndz \: D(z) \: \sbessel_\el[k \eta(z)]
\end{equation}
and
\begin{equation}
F_\el^{\tplab}(k) = \int \dx z \: \dgdz \: \sbessel_\el[k \eta(z)]
\spcend .
\end{equation}
The integration required to compute $F_\el^\tplab(k)$ is performed over $z$ from zero to the epoch of recombination, whereas, in practice, the integration range for $F_\el^\ndlab(k)$ is defined by the source redshift distribution function \dndz.  The function $D(z)$ is the linear growth factor for the matter distribution (calculated from \cmbfast\footnote{\url{http://www.cmbfast.org/}} \citep{seljak:1996} by computing the transfer function for different redshifts), $g \equiv (1+z)D(z)$ is the linear growth suppression factor, $\sbessel_\el[k\eta(z)]$ is the spherical Bessel function and $\eta(z)$ is the conformal look-back time.  The bias factor $b$ is assumed to be redshift independent.

For the evolution of $D(z)$ we consider both the standard model dominated by a cosmological constant and also alternative models dominated by a time varying dark energy with negative pressure, $\rho = \rho_0 (1+z)^{3(1+\w)}$, and energy density that is spatially inhomogeneous.  These alternative models are parameterised by the equation-of-state parameter $\w$, defined as the ratio of pressure to density.  The standard inflationary model assumes a homogeneous field (the cosmological constant) with $\w=-1$, however in general dark energy models are characterised by values of $\w<0$.  For instance, topological defects can be phenomenologically represented by an equation-of-state parameter $-2/3 \leq \w \leq -1/3$ (\eg\ \citealt{friedland:2003}), quintessence models imply $-1 < \w < 0$ \citep{wetterich:1988,caldwell:1998} and phantom models have equation-of-state parameter $\w<-1$ (these last models, however, violate the null dominant energy condition; see \citet{carroll:2003} for a detailed discussion).

A convincing explanation for the origin and nature of the dark energy within the framework of particle physics is lacking, nevertheless in general the models considered in the literature produce a \w\ that varies with time.  For most dark energy models, however, the equation-of-state changes slowly with time and a standard approximation is that (at least during a given epoch) \w\ can be considered as a constant equation-of-state parameter \citep{wang:2000}.  Henceforth, we consider \w\ to be constant as a useful approach to extract fundamental properties of the dark energy.

\subsection{Wavelet covariance estimator}
\label{sec:wav_cov}

The covariance of the wavelet coefficients is used as an estimator to detect any cross-correlation between the \cmb\ and the galaxy density distribution.  A positive covariance indicates a positive cross-correlation between the data.
The wavelet coefficient covariance estimator is defined as the sum over all points on the wavelet domain sky, of the product of the wavelet coefficient maps:
\begin{equation}
\label{eqn:wcov}
\wcovest^{\ndlab\tplab}(\scaleab, \eulc) = \frac{1}{N_{\eula\eulb}}
\sum_{\eula,\eulb} \:
\weight_{\eula\eulb} \:
\skywav_\wav^\ndlab(\scaleab,\euls) \:
\skywav_\wav^\tplab(\scaleab,\euls)
\spcend ,
\end{equation}
where $N_{\eula\eulb}$ is the number of samples in the wavelet domain sky, $\weight_{\eula\eulb}$ is a weighting function and $\skywav_\wav^\ndlab(\scaleab,\euls)$ and $\skywav_\wav^\tplab(\scaleab,\euls)$ are the wavelet coefficients of the galaxy density distribution and \cmb\ respectively.  
We choose weights to reflect the relative size of the pixels on the sky in the wavelet domain (equiangular sampling is used in the wavelet coefficient Euler angle domain), where
\mbox{$\weight_{\eula\eulb}= \frac{\pi}{2} \sin\eulb $}. 
This weighting scheme ensures regions near the poles of the coordinate system do not have a greater influence on the estimated covariance than regions near the equator.

We may also average the covariance estimator over orientations, so that we obtain an overall covariance measure for the given scales:
\begin{equation}
\wcovest^{\ndlab\tplab}(\scaleab) = \frac{1}{N_{\eulc}}
\sum_\eulc \:
\wcovest^{\ndlab\tplab}(\scaleab, \eulc)
\spcend ,
\end{equation}
where $N_{\eulc}$ is the number of samples in the wavelet domain orientational component.  This measure is still sensitive to directional structure when using a directional spherical wavelet, just as
$\wcovest^{\ndlab\tplab}(\scaleab, \eulc)$ and
$\wcovest^{\ndlab\tplab}(\scaleab)$
are both sensitive to localised spatial structure in $(\eula,\eulb)$.

A theoretical prediction of the wavelet covariance may be specified for a given cosmological model.  \citet{vielva:2005} derive this for azimuthally symmetric spherical wavelets, however the extension to directional wavelets is non-trivial.  We present the derivation of the theoretical wavelet covariance for directional wavelets in \appn{\ref{appn:theo_covar}}, stating here the expression for the theoretical covariance obtained:
\begin{equation}
\label{eqn:theoxcorr}
\wcov^{\ndlab\tplab}(\scaleab, \eulc) =
\sum_{\el=0}^\infty \:
\pixw{}^2 \:
\bm{\ndlab} \:
\bm{\tplab} \:
\clnttheo
\sum_{\m=-\el}^\el \left| (\wav_{\scaleab})_\elm \right|^2
\spcend ,
\end{equation}
where $\wav_\elm$ are the spherical harmonic coefficients of the wavelet.  In practice at least one of the functions in \eqn{\ref{eqn:theoxcorr}} has a finite band limit so that negligible power is present in those coefficients above a certain \elmax.  All summations over \el, here and subsequently, may therefore be truncated to \elmax.

\subsection{Comparison of wavelet covariance estimators}

We compare the expected performance of various spherical wavelets for detecting the \isw\ effect in this section.
\citet{vielva:2005} perform a similar analysis to compare the performance of the \smhw\ estimator to real and harmonic space estimators, showing the effectiveness of the \smhw\ estimator.
Instead we focus here on comparing wavelets; thus we extend the analysis to directional wavelets.

The expected signal-to-noise ratio (\snr) of the wavelet covariance estimator 
of the \cmb\ and the \lss\ density distribution may be used to compare the expected performance of various wavelets for detecting the \isw\ effect.
The expected \snr\ for a particular scale is given by the ratio of the expected value of the wavelet covariance estimator and its dispersion:
\begin{equation}
\snr_\wav(\scaleab) =
\frac{\opnexpv \wcovest^{\ndlab\tplab}(\scaleab) \clsexpv}
{\Delta \wcovest^{\ndlab\tplab}(\scaleab)}
\spcend , 
\end{equation}
where for a directional wavelet the variance of the wavelet covariance estimator is given by
\begin{eqnarray}
\left[ \Delta \wcovest^{\ndlab\tplab}(\scaleab) \right]^2 &=&
\opnexpv \left[\wcovest^{\ndlab\tplab}(\scaleab)\right]^2 \clsexpv
-
\opnexpv \wcovest^{\ndlab\tplab}(\scaleab) \clsexpv^2 \nonumber \\
&=&
\displaystyle
\sum_{\el=0}^\infty \frac{1}{2\el+1} \:
\pixw{}^4 \:
(\bm{\ndlab})^2 \:
(\bm{\tplab})^2\:
\nonumber \\
&& \times  \:
\left( \sum_{\m=-\el}^\el \left| (\wav_{\scaleab})_\elm \right|^2 \right)^2
\left( (\clnttheo)^2 + \cltttheo \clnntheo \right)
\, ,
\label{eqn:wcov_disp}
\end{eqnarray}
where $\cltttheo$ and $\clnntheo$ are the \cmb\ and galaxy count power spectra respectively.
For the case of azimuthally symmetric wavelets \eqn{\ref{eqn:wcov_disp}} reduces to the form given by \citet{vielva:2005}.\footnote{Note that \citet{vielva:2005} use Legendre coefficients to represent the symmetric \smhw, rather than spherical harmonic coefficients.}

The expected \snr\ is computed for the spherical wavelets we consider for a range of dilations (the dilations considered are those that are subsequently used to attempt to detect the \isw\ effect, as defined in \tbl{\ref{tbl:scales}}).
The theoretical power spectra used in this experiment are computed from the cosmological concordance model parameters specified in \tbl{1} of \citet{spergel:2003}
($\Omega_\Lambda=0.71$, $\Omega_{\rm m}=0.29$, $\Omega_{\rm b}=0.047$,
$H_0=72$, $\tau=0.166$, $n=0.99$).  \cmbfast\ is used to simulate the \cmb\ spectrum \cltttheo, the cross-power \clnttheo\ spectrum is simulated in accordance with the theory outlined in \sectn{\ref{sec:theo_crossspec}}, while the actual \nvss\ data is used to provide the \lss\ angular power spectrum \clnntheo.
Ideal noise-free and full-sky conditions are assumed for this experiment.  Beam and pixel windowing are not considered in the results presented, although when included these factors make only a very minimal difference to the numerical values obtained.  The expected \snr\ computed for each wavelet for a range of scales is illustrated in \fig{\ref{fig:snr}}.
The relative expected performance of the \smw\ for detecting the \isw\ effect is low, thus we do not consider this wavelet any further.  The \smhw\ and the \sbw\ are of comparable expected performance.  However, notice that the elliptical \smhw\ (illustrated by the off-diagonal values of \fig{\ref{fig:snr}~(b)}) and the \sbw\ are slightly superior to the azimuthally symmetric \smhw\ (illustrated by the diagonal values of \fig{\ref{fig:snr}~(b)}).
It may appear counter intuitive initially that asymmetric wavelets may perform better at detecting the \isw\ effect when the data are isotropic fields.  Although the statistics of the fields are globally isotropic, this does not preclude local oriented features in the data.  Indeed,
the distribution of the ellipticity of peaks in a isotropic Gaussian random field has been derived by \citet{barreiro:1997,barreiro:2001}, illustrating that one would expect to see local rotationally invariant features.
The maximum expected \snr\ achievable with both the \smhw\ and \sbw\ is approximately 4, compared to an estimated best achievable \snr\ of approximately 7.5 for a perfect survey \citep{afshordi:2004}.

\newlength{\snrplotwidth}
\setlength{\snrplotwidth}{70mm}

\begin{figure}
\centering
\subfigure[All wavelets for isotropic dilations]{\includegraphics[clip=,width=\snrplotwidth]{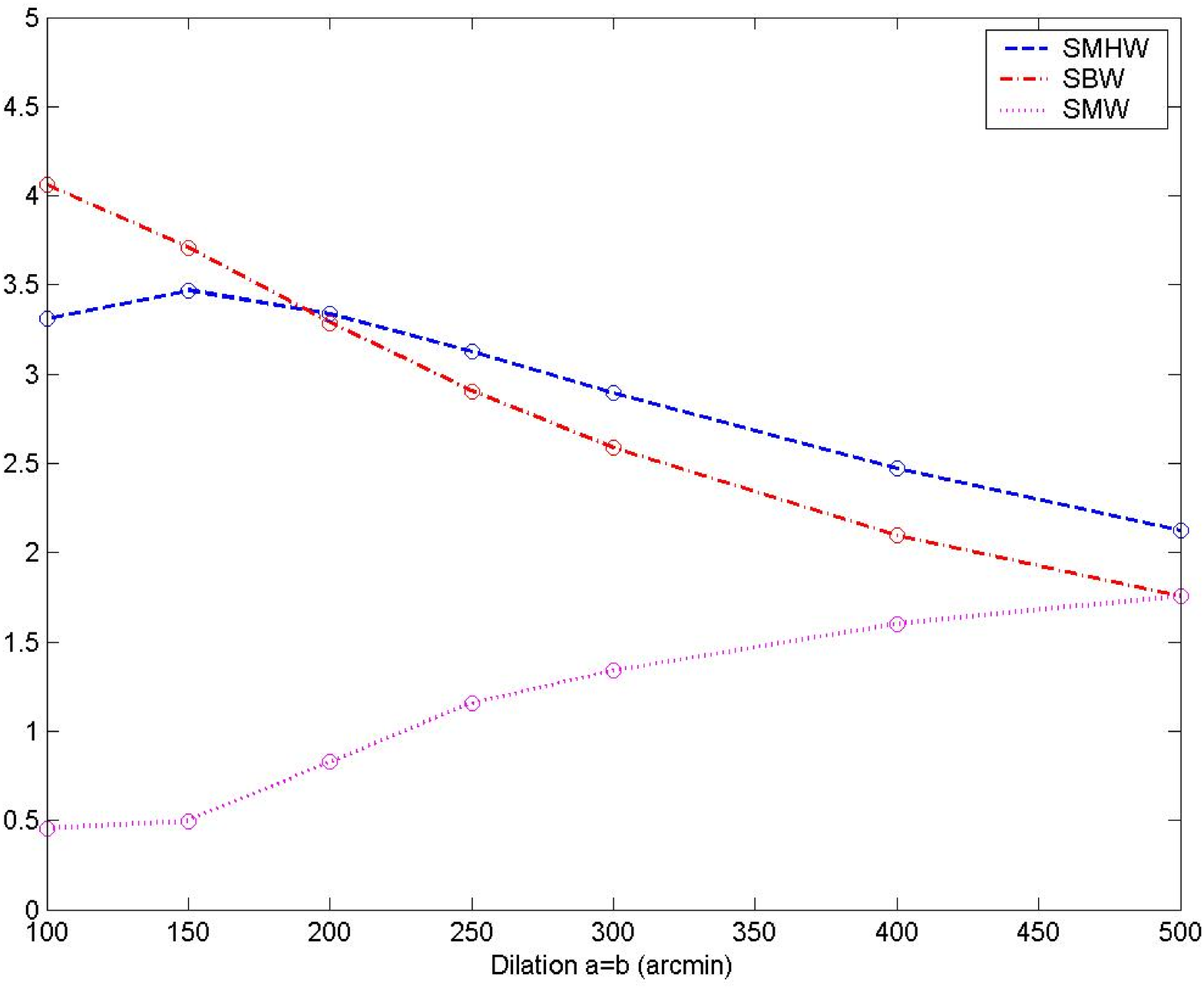}}
\subfigure[\smhw\ for all dilations]{\includegraphics[clip=,width=\snrplotwidth]{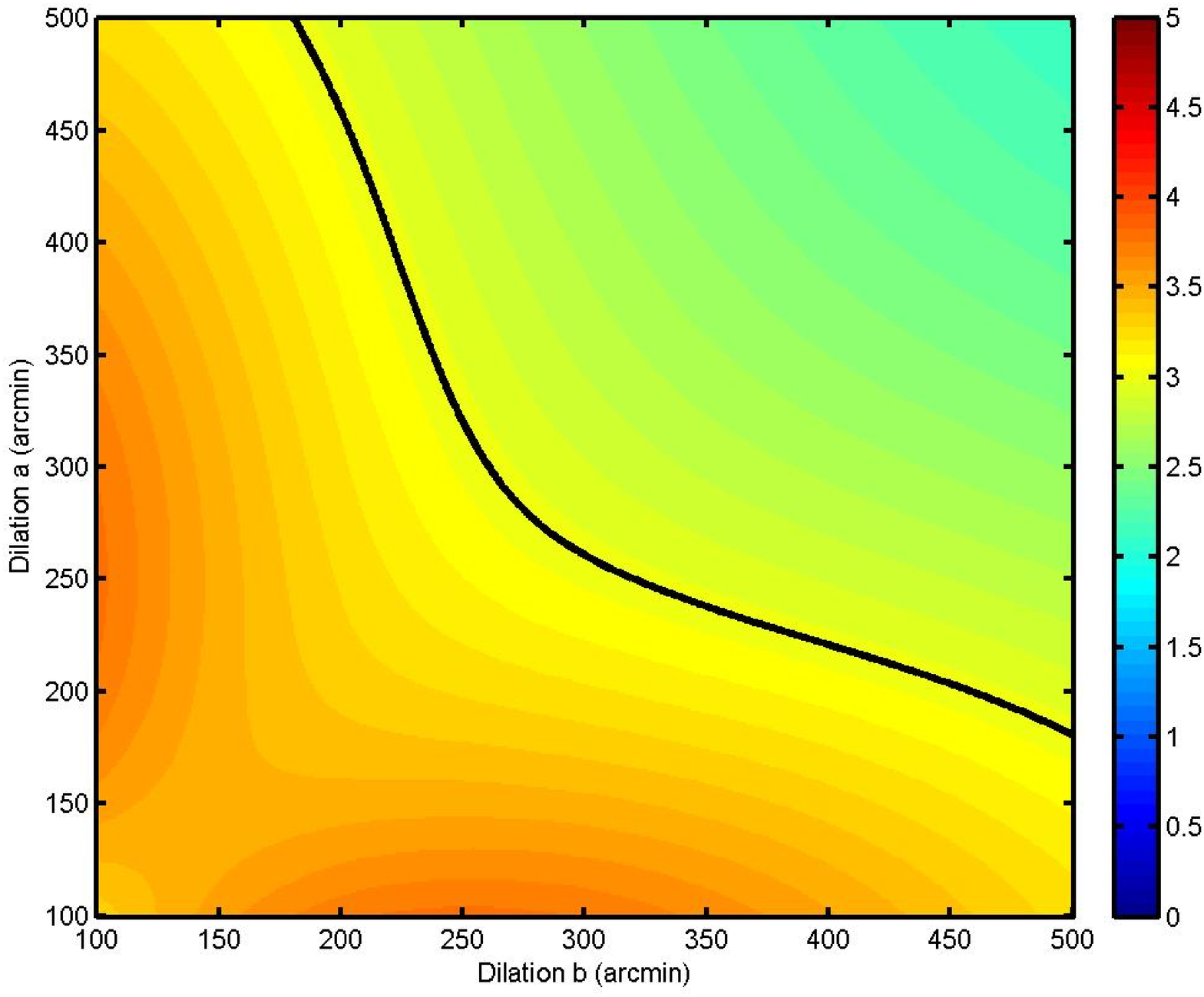}}
\subfigure[\sbw\ for all dilations]{\includegraphics[clip=,width=\snrplotwidth]{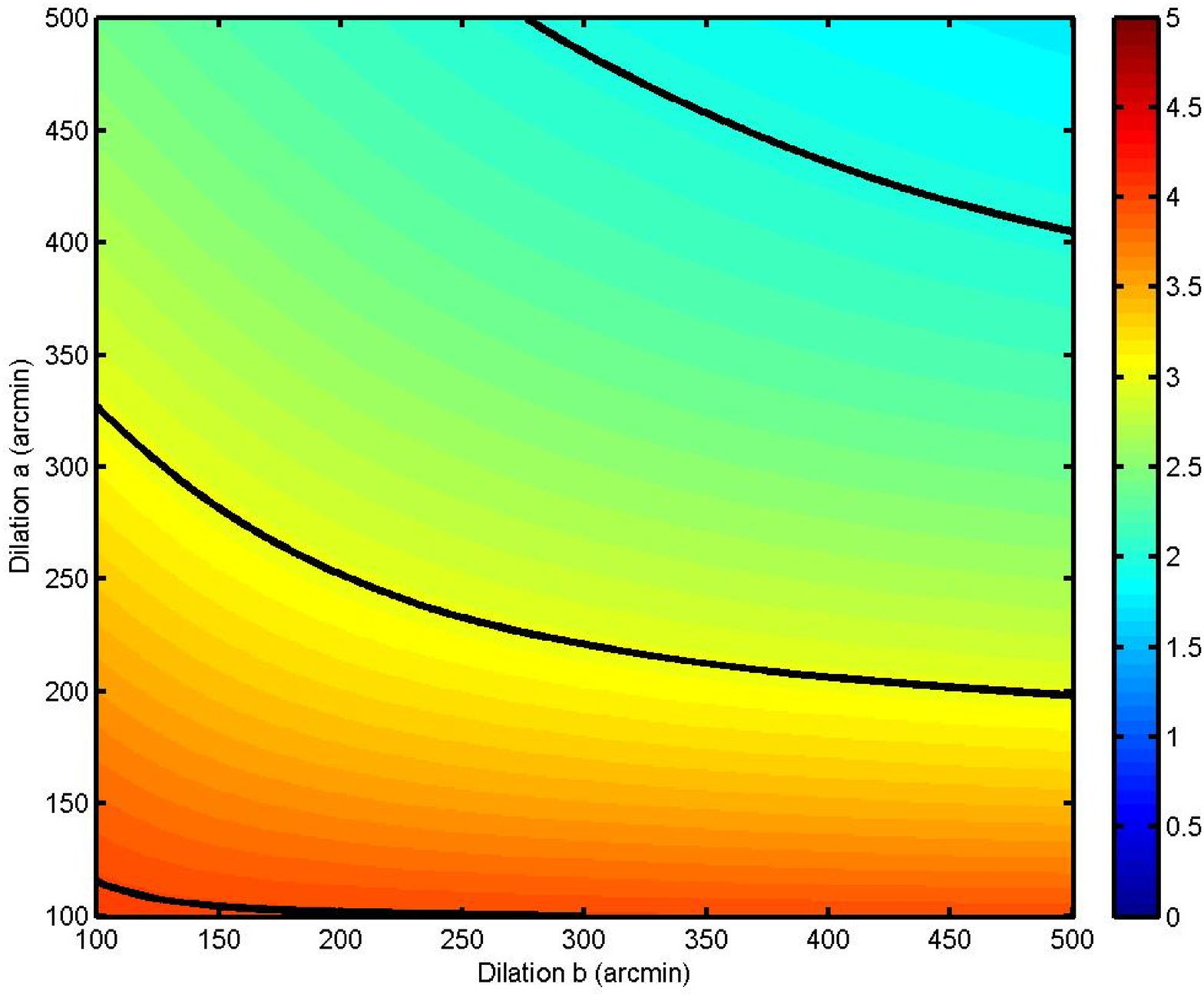}}
\caption{Expected \snr\ of the wavelet covariance estimator of \cmb\ and radio source maps (cosmological parameters are chosen according to the concordance model).  All spherical wavelets are considered for isotropic dilations in panel (a): \smhw\ (blue, dashed); \sbw\ (red, dotted-dashed); \smw\ (magenta, dotted).  Due to the relatively poor expected performance of the \smw, the \snr\ for all anisotropic dilations in only shown for the \smhw\ and \sbw\ in panels (b) and (c) respectively.  Contours at \snr\ values of two, three and four are also plotted in panels (b) and (c).
  }
\label{fig:snr}
\end{figure}


\section{Analysis procedure}
\label{sec:analysis}

We describe in this section the data and an overview of the analysis procedure used to attempt to detect the \isw\ effect using directional spherical wavelets.

\newlength{\mapplotwidth}
\setlength{\mapplotwidth}{75mm}

\begin{figure}
\centering
\subfigure[\wmap]{\includegraphics[clip=,width=\mapplotwidth]{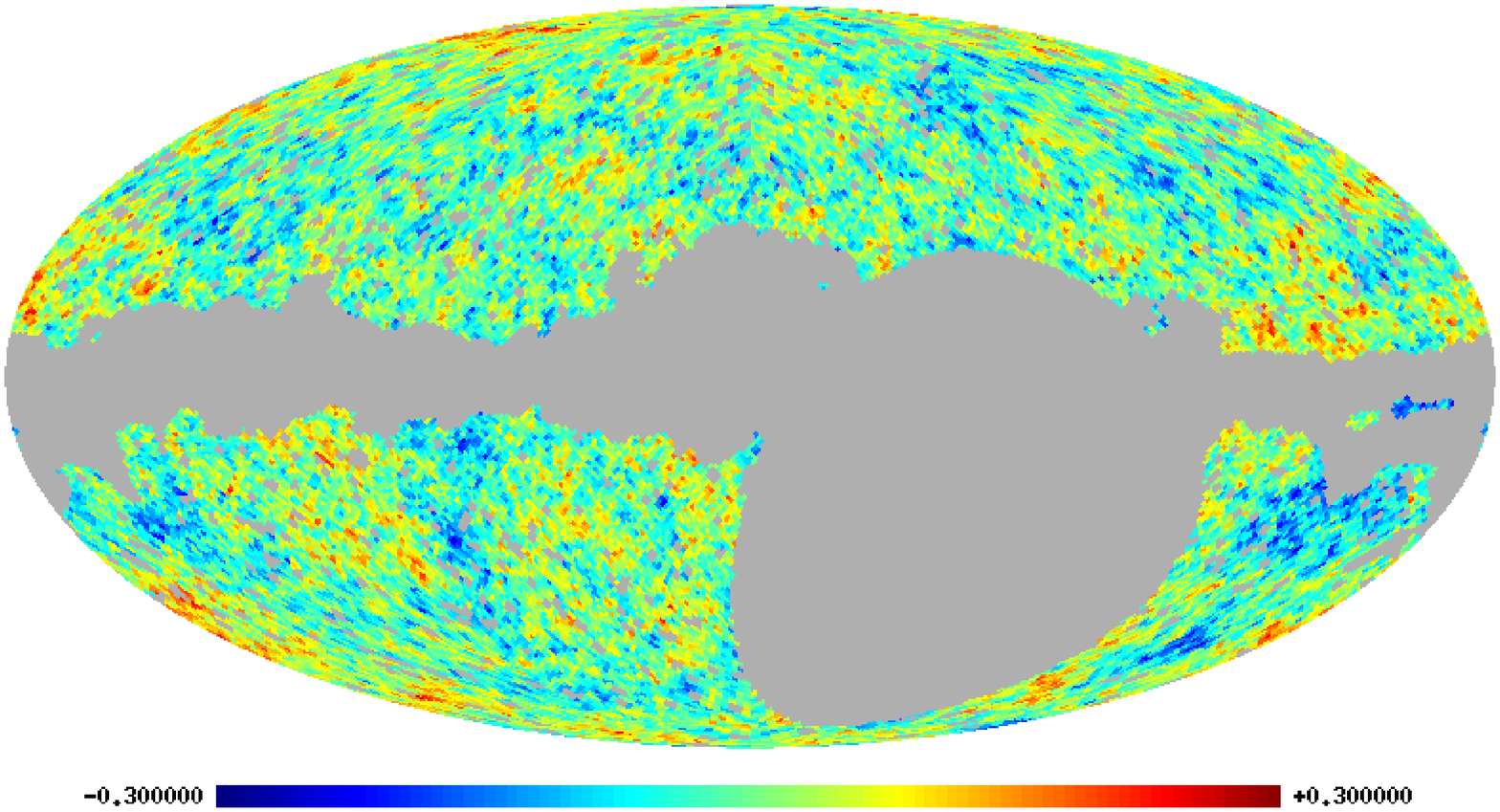}}
\subfigure[\nvss]{\includegraphics[clip=,width=\mapplotwidth]{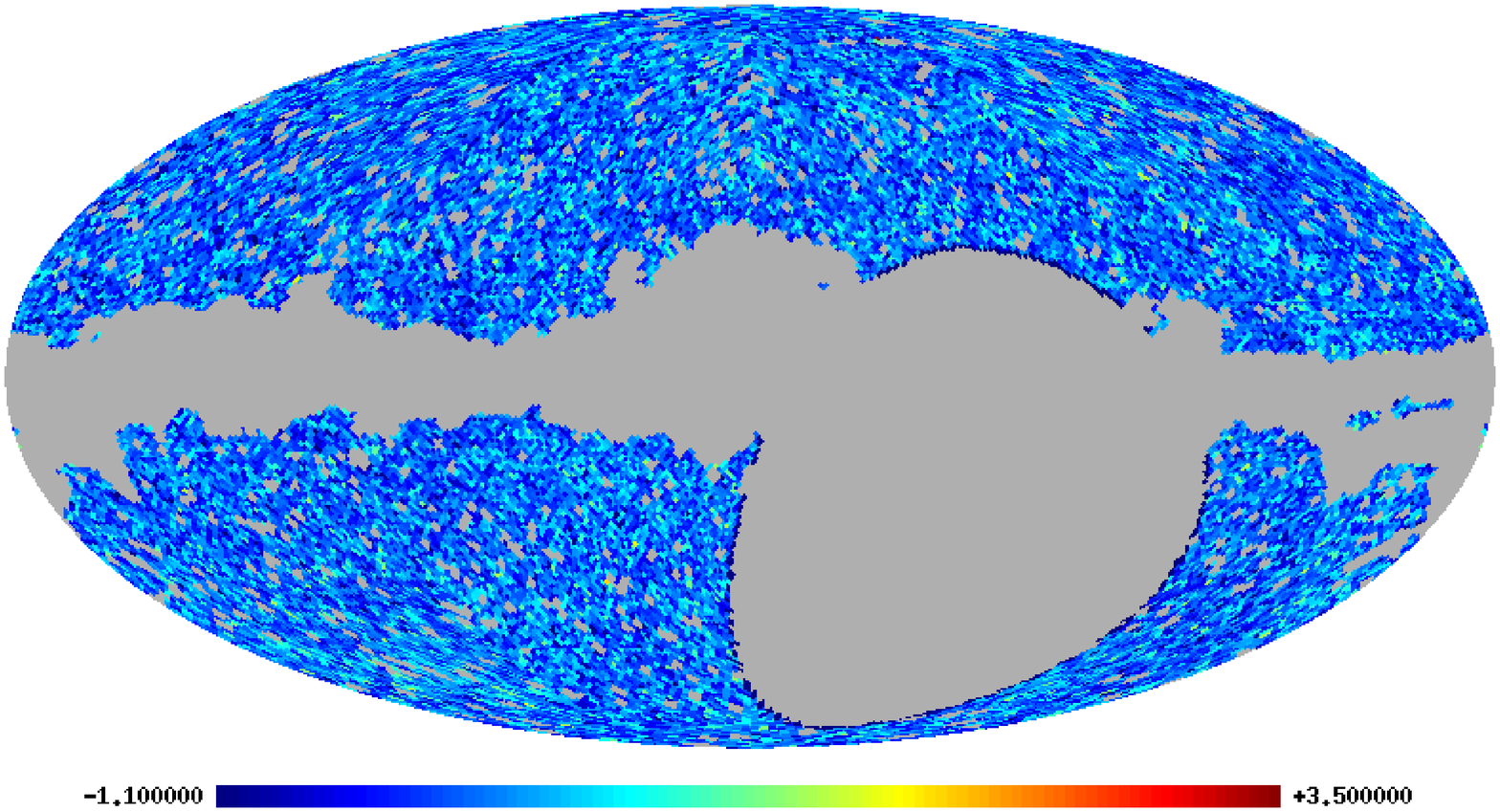}}
\caption{\wmap\ and \nvss\ maps after application of the joint mask.  The maps are represented in the \healpix\ format at a resolution of $\nside=64$ (corresponding to a pixel size of $\sim\!\!55\arcmin$).}
\label{fig:maps}
\end{figure}

\subsection{WMAP data}
\label{sec:wmap_data}

To minimise the contribution of foregrounds and systematics to \cmb\ anisotropy measurements the \wmap\ assembly contains a number of receivers that observe at a range of frequencies.
The \wmap\ team and other independent groups have proposed various constructions of \cmb\ maps from data measured by different receivers and bands in order to minimise foreground contributions
(template based methods: \citealt{bennett:2003b}; template independent methods: \citealt{bennett:2003b}, \citealt{tegmark:2003}, \citealt{eriksen:2004}).
In this work, we use the template based foreground removed maps \citep{bennett:2003b} to construct the co-added map proposed by the \wmap\ team
and used in their non-Gaussianity analysis \citep{komatsu:2003}.
Following the data processing pipeline specified by \citet{komatsu:2003}, the foreground cleaned \wmap\ maps for which the \cmb\ is the dominant signal (two Q-band maps at 40.7\ghz, two V-band maps at 60.8\ghz\ and four W-band
maps at 93.5\ghz) are combined to give the signal-to-noise ratio enhanced co-added map.
The conservative \kpzero\ exclusion mask provided by the \wmap\ team is applied to remove remaining Galactic emission and bright point sources.
The foreground removed maps used to construct the co-added map and the \kpzero\ mask
are available from the
Legacy Archive for Microwave Background Data Analysis (\lambdaarch).\footnote{\url{http://cmbdata.gsfc.nasa.gov/}}
The maps are provided in the \healpix\footnote{\url{http://healpix.jpl.nasa.gov/}}
\citep{gorski:2005} format at a resolution of $\nside=512$ (giving $12 \nside{}^2 \sim 3\times10^6$ pixels on the sphere).
Since the \isw\ signal we hope to detect is expected to be generated by structure on scales greater than $2^\circ$ \citep{afshordi:2004}, we down-sample the co-added map constructed to $\nside=64$ ($\sim 5\times10^4$ pixels).  This corresponds to a pixel size of approximately $55\arcmin$ and so should be of sufficient resolution to detect the \isw\ signal.
The co-added \wmap\ map used in the subsequent analysis is displayed in \fig{\ref{fig:maps}~(a)} (with the joint \kpzero-\nvss\ mask applied, as described in \sectn{\ref{sec:nvss_data}}).

\subsection{NVSS data}
\label{sec:nvss_data}

We use the \nvss\ radio source catalogue as a local tracer of the \lss.
The catalogue covers approximately 80\% of the sky and contains measurements of almost $2\times10^6$ point sources with a minimum flux density of $\sim\!\!2.5$mJy.  Although the distance of individual sources is largely unknown, the sources are thought to be distributed in the redshift range $0<z<2$ with a peak in the distribution at $z\sim0.8$ \citep{boughn:2002}.  The \isw\ signal we hope to detect is expected to be produced at $z\sim0.4$ with negligible contribution from $z>1.5$ \citep{afshordi:2004}.
Furthermore, the correlations induced by the \isw\ effect are a relatively large scale phenomenon, hence any detection will be cosmic variance limited.
The near full-sky coverage and source distribution of the \nvss\ data thus make it an ideal probe of the local matter distribution to use when searching for the \isw\ effect.  Indeed, the \nvss\ catalogue has been used to make positive detections of the \isw\ effect already using the real space correlation function \citep{boughn:2002,boughn:2004,nolta:2004}.

As first noticed by \citet{boughn:2002}, the \nvss\ catalogue mean source density varies with declination.  We correct this systematic by scaling each iso-latitude band to impose a constant mean density.  The iso-latitude bands are defined by the continuous root-mean-squared ({RMS}) noise regions of the survey (see \fig{10} of \citealt{condon:1998}).
Alternative correction strategies were also considered (\eg\ the one proposed by \citealt{nolta:2004}), however we found cross-correlation detections to be insensitive to the particular correction procedure.

We represent the corrected \nvss\ source distribution in the \healpix\ format at the same resolution of the \wmap\ data (\mbox{$\nside=64$}).  We consider only the sources in the \nvss\ catalogue above 2.5mJy, corresponding to an identification completeness of 50\% \citep{condon:1998}.
No observations are made in the \nvss\ catalogue for equatorial declination 
\eqdec\ lower than $-50^\circ$
and the coverage in the range $-50^\circ < \eqdec < -37^\circ$ is not sufficient.  Hence, we consider only sources with an equatorial declination $\eqdec \geq -37^\circ$.  
With these constraints we obtain a galaxy distribution map containing $\sim\!\! 1.6 \times 10^6$ sources with an average number of $40.4$ counts per pixel.
Since the \wmap\ mask and \nvss\ coverage each exclude various sections of the sky, we construct a joint mask 
(\kpzero$\:+\: \eqdec\!<\!-37^\circ$) 
that leaves only those common pixels that remain in both maps.  This mask is applied to both the co-added \wmap\ and \nvss\ maps used in the subsequent analysis.  The \nvss\ map with the joint map applied is displayed in \fig{\ref{fig:maps}~(b)}.

\subsection{Simulations}

Monte Carlo simulations are performed to construct significance measures for
the wavelet covariance statistics used to detect the \isw\ effect.
1000 Gaussian simulations of the \wmap\ data are constructed from the power spectrum produced by \cmbfast\ using the cosmological concordance model parameters specified in \tbl{1} of \citet{spergel:2003}.
For each realisation we simulate the \wmap\ observing strategy and then construct a simulated co-added map.
Measurements made by the Q-, V- and W-band receivers are simulated by convolving with realistic beams and adding anisotropic \wmap\ noise for each receiver.  The procedure described in \sectn{\ref{sec:wmap_data}} to construct the co-added map is then performed on the simulated data.  The simulated co-added map is then down-sampled to $\nside=64$ and the joint mask is applied.

\subsection{Procedure}

The analysis procedure consists of computing the wavelet covariance estimator described in \sectn{\ref{sec:wav_cov}} from the wavelet coefficients of the co-added \wmap\ and \nvss\ data computed for a range of scales and, for the directional wavelets, a range of \eulc\ orientations.
We consider only those scales where the \isw\ signal is expected to be significant \citep{afshordi:2004}: the wavelet scales considered and the corresponding effective size on the sky of the wavelets are shown in \tbl{\ref{tbl:scales}}.
Each of the directional wavelets considered are rotationally invariant under integer azimuthal rotations of $\pi$, thus the azimuthal rotation angle \eulc\ effectively lies in the domain $[0,\pi)$.
For directional wavelets we consider five evenly spaced \eulc\ orientations in the domain $[0,\pi)$.
Any deviation from zero in the wavelet covariance estimator for any particular scale or orientation is an indication of a correlation between the \wmap\ and \nvss\ data and hence a possible detection of the \isw\ effect.  An identical analysis is performed using the simulated co-added \cmb\ maps in place of the \wmap\ data in order to construct significance measures for any detections made.
Finally, we use any detections of the \isw\ effect to constrain dark energy parameters.


The application of the joint \kpzero-\nvss\ mask distorts wavelet coefficients corresponding to wavelets with support that overlaps the mask exclusion region.  These contaminated wavelet coefficients must be removed from the wavelet covariance estimate.  We construct an extended coefficient mask for each scale to remove all contaminated wavelet coefficients from the analysis.
Binary morphological operations, commonly applied on the plane in image processing \citep{gonzalez:2003}, are extended to the sphere and applied to construct the extended coefficient mask.  A morphological dilation adds pixels to the boundaries of regions, whereas a morphological erosion removes pixels from boundaries.  Firstly, the original joint mask is `opened' (a morphological dilation followed by an erosion) to remove point source regions whilst maintaining the size of the original central masked region.
The central masked region is then extended by half the effective size of the wavelet (at the particular scale) by performing a morphological erosion.
The original point source mask regions are then replaced by applying the original mask (point source regions are removed initially to ensure that they are not expanded when performing the final erosion).
The result of this procedure is to extend the central region of the joint mask by half the effective size of the wavelet, while maintaining the point source regions.  This ensures that all coefficients corresponding to wavelets that overlap with the masked Galactic plane are excluded from the analysis.  However, wavelets that overlap with masked point sources are not removed.  
Retaining these wavelet coefficients induces minimal distortion due to the large support of the wavelets relative to the size of the masked point sources.
This is a less conservative procedure for constructing extended coefficient masks than used in our non-Gaussianity analysis \citep{mcewen:2005b}, however the low resolution of the data and the large initial joint mask reduce drastically the number of pixels available initially.  If we construct more conservative extended masks we find that we quickly and significantly reduce the number of pixels considered and thus the efficiency of the analysis.  To ensure any correlations detected are not due to contamination that is not excluded by the extended mask we have also performed a number of tests using masks with larger exclusion regions.  We find that the correlation signals that we subsequently detect (see \sectn{\ref{sec:results}}) remain when using these larger exclusion masks.  Hence our choice of mask appears appropriate as it is more efficient than more conservative masks but still removes the contaminated wavelet coefficients.

\begin{table*}
\begin{minipage}{120mm}
\centering
\caption{Wavelet scales considered in attempting to detect the \isw\ effect.  The effective sizes on the sky of the various wavelets for each wavelet scale are also shown.  Note that for anisotropic dilations the wavelets have two effective sizes on the sky defined in orthogonal directions (see text).}
\label{tbl:scales}
\begin{tabular}{lcccccccccccc} \hline
Scale & 1 & 2 & 3 & 4 & 5 & 6 & 7 \\ \hline
Dilation \scalea  & 100\arcmin & 150\arcmin & 200\arcmin & 250\arcmin & 300\arcmin & 400\arcmin & 500\arcmin \\
 Size on sky $\effsize^{\smhw,\, \sbw}_{1,2}(\scalea)=\effsize^\smw_1(\scalea)$ & 282\arcmin & 424\arcmin & 565\arcmin & 706\arcmin & 847\arcmin & 1130\arcmin & 1410\arcmin \\
      Size on sky $\effsize^\smw_2(\scalea)$ & 31.4\arcmin & 47.1\arcmin & 62.8\arcmin & 78.5\arcmin & 94.2\arcmin & 126\arcmin & 157\arcmin \\
\hline
\end{tabular}
\end{minipage}
\end{table*}

\section{Results and discussion}
\label{sec:results}

In this section we describe the results obtained when performing the directional spherical wavelet analysis procedure described previously.  We describe the detections of the \isw\ effect that we make, before using the detections to constrain dark energy parameters.

\subsection{Detection of the ISW effect}

Firstly, we describe the detection of positive correlations between the \wmap\ and \nvss\ data using various wavelets.   To examine the source of the correlation detected we perform some preliminary tests to ascertain whether perhaps foregrounds or systematics are responsible.  We then localise regions on the sky that contribute most strongly to the correlation signal and examine these in more detail.

\subsubsection{Detections}

The wavelet covariance of the \wmap-\nvss\ data are shown in \fig{\ref{fig:xcorr}} for each of the wavelets considered.  The covariance values are shown in units of \nsigma, the number of standard deviations that the data measurement differs from the mean of 1000 Monte Carlo simulations for each covariance statistic.  
On examining the distribution of the wavelet covariance statistics of the simulations, the covariance statistics themselves appear to  approximately Gaussian distributed.  This implies that the approximate significance of any detection of a non-zero covariance can be inferred directly from the \nsigma\ level.\footnote{Ideally one would compute the significance of a detection directly from the Monte Carlo simulations.  Indeed we do this to construct significance levels in \fig{\ref{fig:xcorr}}, however for 1000 simulations the maximum significance one can measure is 99.9\%.  For many cases this does not give a fine enough resolution, thus we use the \nsigma\ level to compare the significance of detections made on different scales and with different wavelets.  For computational reasons we cannot extend easily the number of simulations far above 1000.}
In any case, 68\%, 95\% and 99\% significance regions are plotted in \fig{\ref{fig:xcorr}} also (these significance levels are computed from the simulations directly but correspond approximately to 1.00$\sigma$, 1.96$\sigma$ and 2.57$\sigma$ respectively).
The left panel in \fig{\ref{fig:xcorr}} shows the results for the azimuthally symmetric \smhw\ (with isotropic dilations), whereas the other panels correspond to directional wavelets.  These latter panels exhibit a jagged structure due to the ordering of the wavelet scale index the covariance is plotted against.  The wavelet scale index is ordered by sequentially concatenating rows from the 2-dimensional $(\scaleab)$ dilation space.  It would be preferable to view the data as a surface in $(\scaleab)$ space, however it would not be possible to view the data and the six significance level surfaces at once. 
The scale index $j$ used in \fig{\ref{fig:xcorr}} and subsequently may be related to the scale index defined in \tbl{\ref{tbl:scales}} for \scalea\ and \scaleb\ by
$\lfloor (j-1)/7 \rfloor$ and ${\rm mod}(j-1,7)+1$ respectively, where $\lfloor x \rfloor$ is the largest integer less than $x$ and ${\rm mod}(\cdot,n)$ is the modulo $n$ function.
%
For all of the wavelets considered, the wavelet covariance of the data lies outside of the 99\% significance level on certain scales.\footnote{Although the \smw\ was shown to be poor at detecting cross-correlations in the data we ran the analysis regardless.  As expected, the analysis results showed that the wavelet is indeed ineffective at detecting any correlation.}


\newlength{\xcorrplotwidth}
\setlength{\xcorrplotwidth}{55mm}

\begin{figure*}
\begin{minipage}{\textwidth}
\centering
\mbox{
\subfigure[Symmetric \smhw]
  {\includegraphics[clip=,width=\xcorrplotwidth]{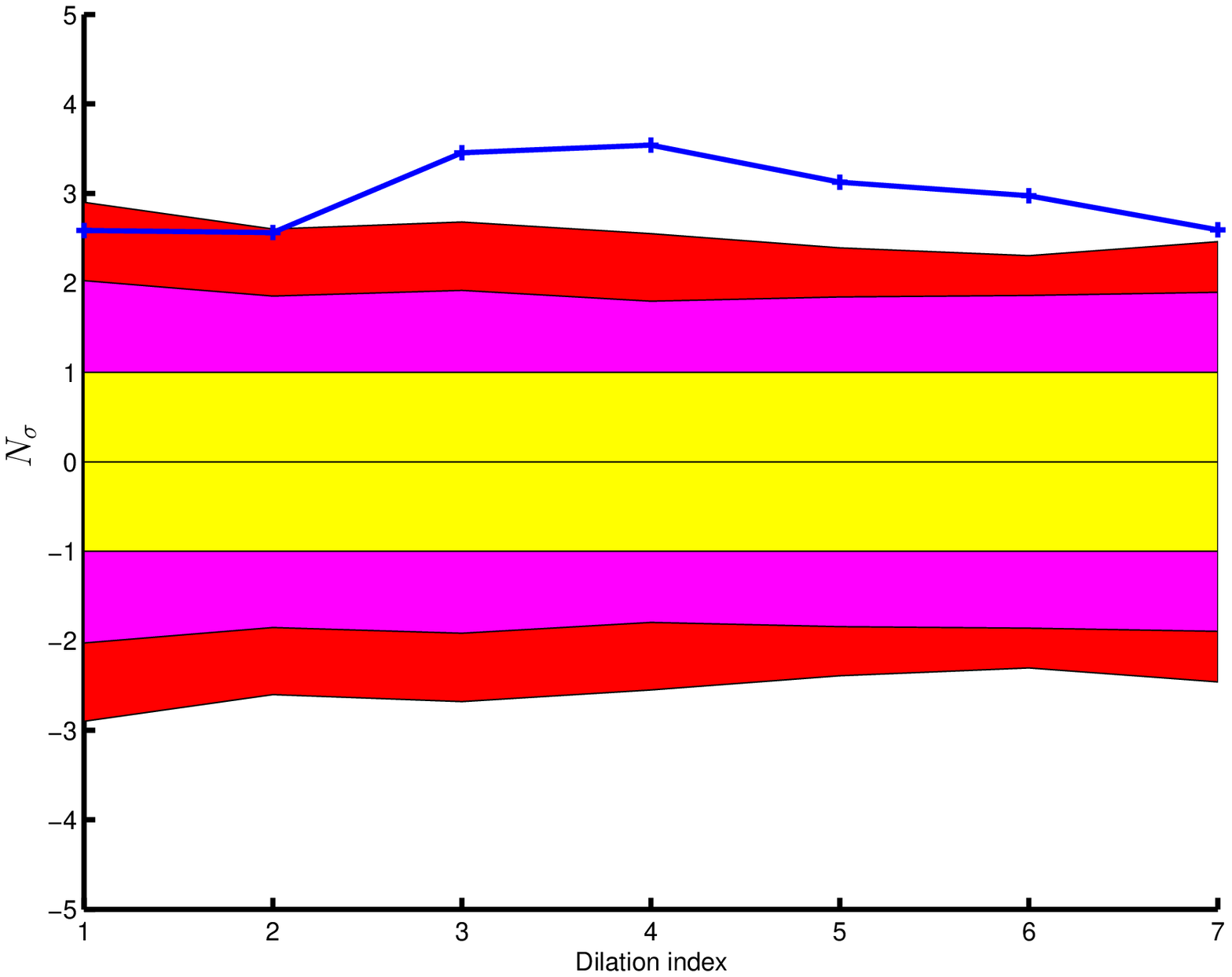}} \quad
\subfigure[Elliptical \smhw]
  {\includegraphics[clip=,width=\xcorrplotwidth]{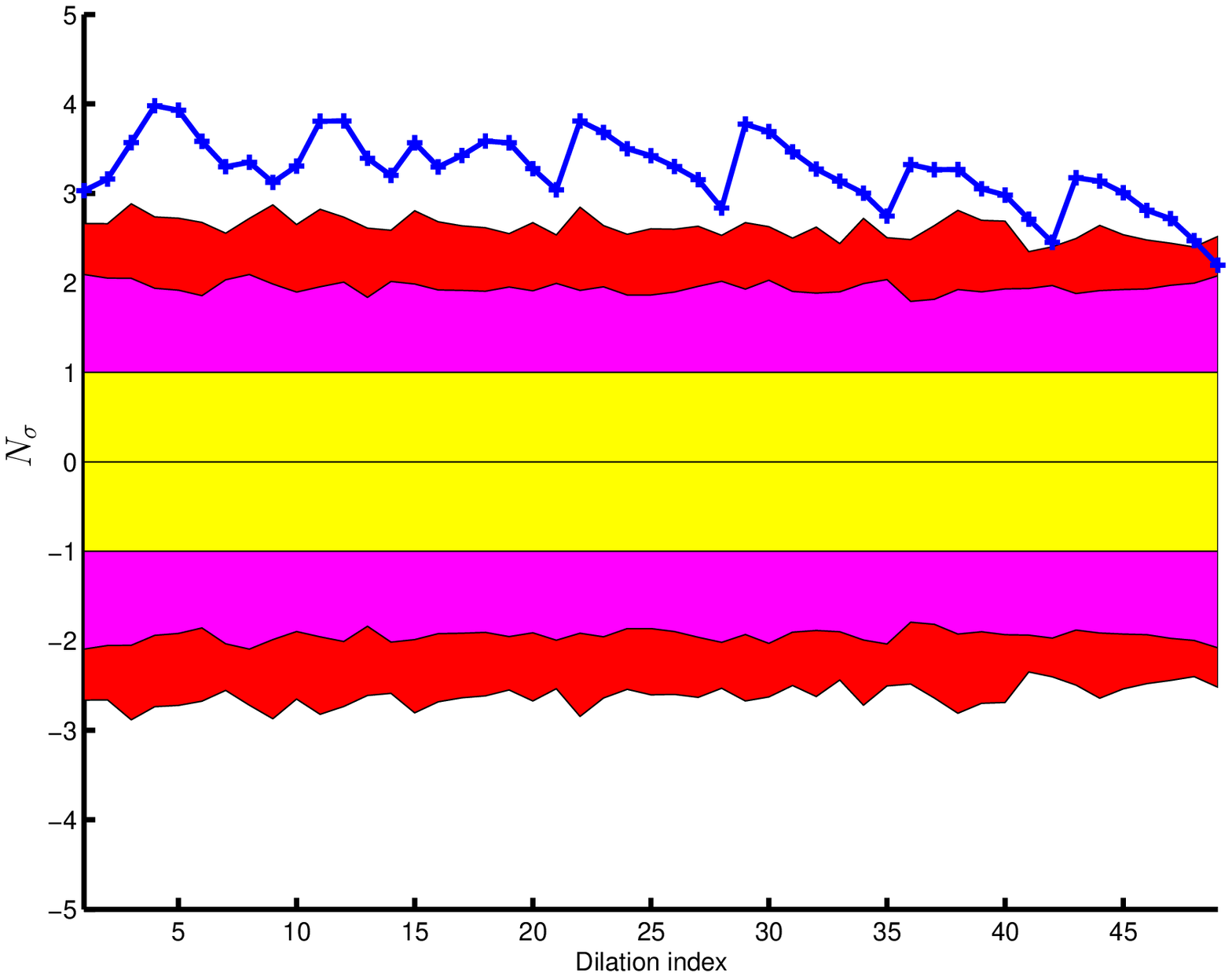}} \quad
\subfigure[\sbw]
  {\includegraphics[clip=,width=\xcorrplotwidth]{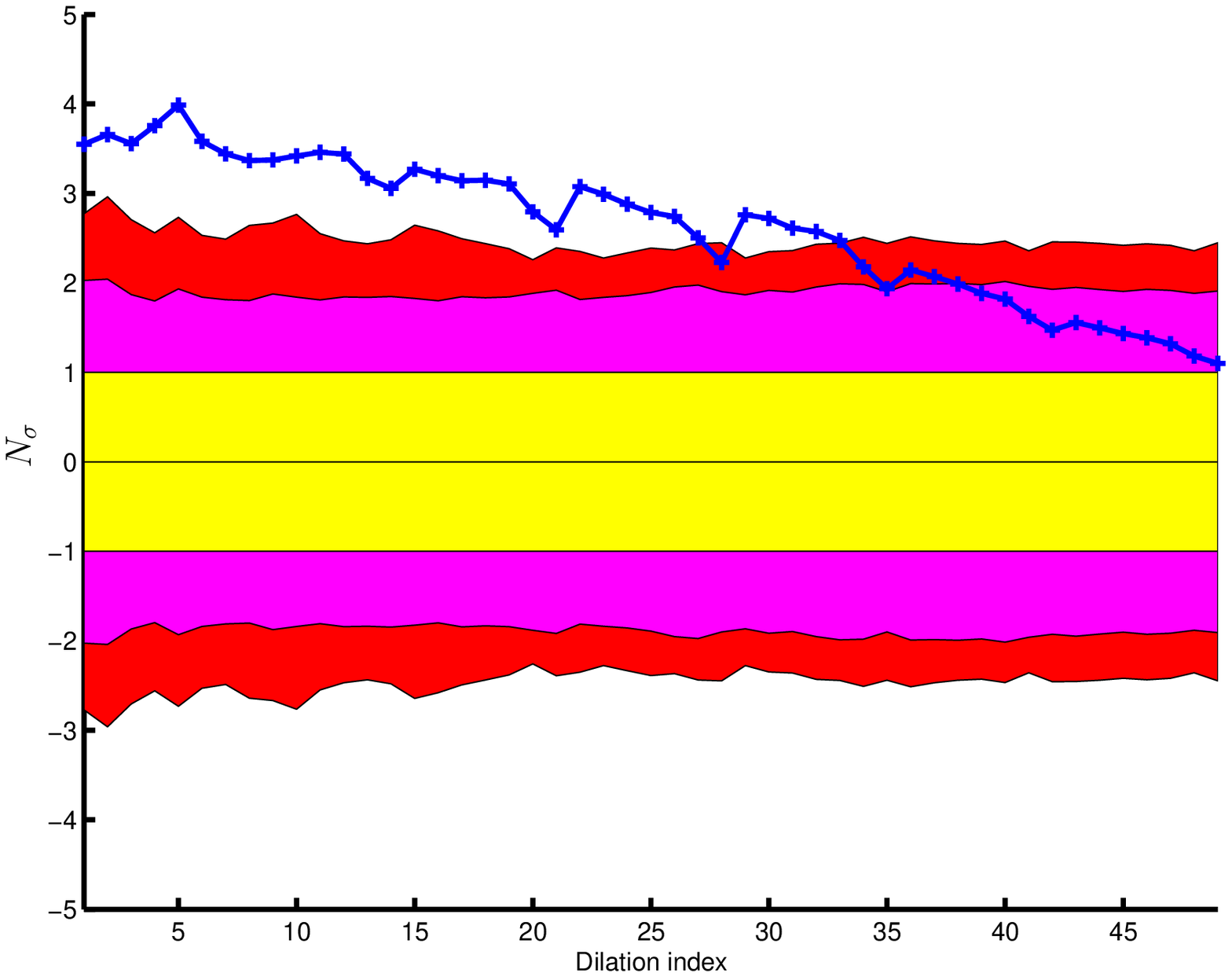}}
}
\caption{Wavelet covariance statistics in units of \nsigma\ for the \wmap\ and \nvss\ data against wavelet scale index.  
Significance levels obtained from 1000 Monte Carlo simulations are shown by the shaded regions for 68\% (yellow/light-grey), 95\% (magenta/grey) and 99\% (red/dark-grey) levels. See text for a discussion of the ordering of the dilation index.}
\label{fig:xcorr}
\end{minipage}
\end{figure*}


In \fig{\ref{fig:nsigma}} we show the \nsigma\ surfaces for each wavelet in $(\scaleab)$ space.
The maximum \nsigma\ of all detections made with the directional \smhw\ and \sbw\ is 3.9 for both wavelets (the maximum detection made with the \sbw\ is slightly greater than that made with the \smhw, however to one decimal place they are equivalent).  Using the
symmetric \smhw\ (\ie\ using only isotropic dilations) however, one obtains a maximum \nsigma\ of 3.6.
It should be noted that the use of the maximum \nsigma\ to characterise the detection is based on an {\it a posteriori} scale selection: the significance of the detection is characterised for a specific scale, rather than the range of scales examined.
It is interesting to note that for both wavelets the \nsigma\ surfaces displayed in \fig{\ref{fig:nsigma}} are perfectly consistent with the expected \snr\ displayed in \fig{\ref{fig:snr}}: the most significant detections are made on scales with high expected \snr.  Note that the \nsigma\ surface for the \smhw\ (\fig{\ref{fig:nsigma}}~(a)) is not perfectly symmetric about the line $\scalea=\scaleb$.  Although the wavelet has identical shape for dilations $(\scalea,\scaleb)$ and $(\scaleb,\scalea)$, the second wavelet is rotated azimuthally by $\pi/2$ relative to the first wavelet.  Since the starting position differs between the two cases the directional analysis of each probes slightly different directions, resulting in an \nsigma\ surface that is not perfectly symmetric about the diagonal.

The symmetric \smhw\ analysis has already been performed by \citet{vielva:2005}.  We have repeated the analysis here as a consistency check for our analysis and also to confirm the previous work.  The wavelet covariance curve shown in \fig{\ref{fig:xcorr}~(a)}
is similar to the curve obtained by \citet{vielva:2005}, however we make a slightly more significant detection at $3.6\sigma$ compared to the detection made by \citet{vielva:2005} at $3.3\sigma$.  After performing numerous tests with different weighting and masking schemes we believe the difference in significance with \citet{vielva:2005} is due to  
the different mask construction techniques adopted, the different pixelisation of the wavelet domain and the slightly different definition of the \smhw\ used (see \citet{martinez:2002} for the definition used by \citet{vielva:2005}).

In summary, we have detected a positive correlation between the \wmap\ and \nvss\ data, as indicated by a positive wavelet covariance, at the $3.9\sigma$ level on wavelet scales
about $(\scalea,\scaleb)=(100\arcmin,300\arcmin)$.  The sign of the correlation detected (positive) and the scale the detections are made at is consistent with an \isw\ signal.  We next perform some preliminary tests to determine whether the correlation we detect is indeed produced by the \isw\ effect or whether other factors are responsible.

\newlength{\nsigmaplotwidth}
\setlength{\nsigmaplotwidth}{70mm}

\begin{figure}
\centering
\subfigure[\smhw]
  {\includegraphics[clip=,width=\nsigmaplotwidth]{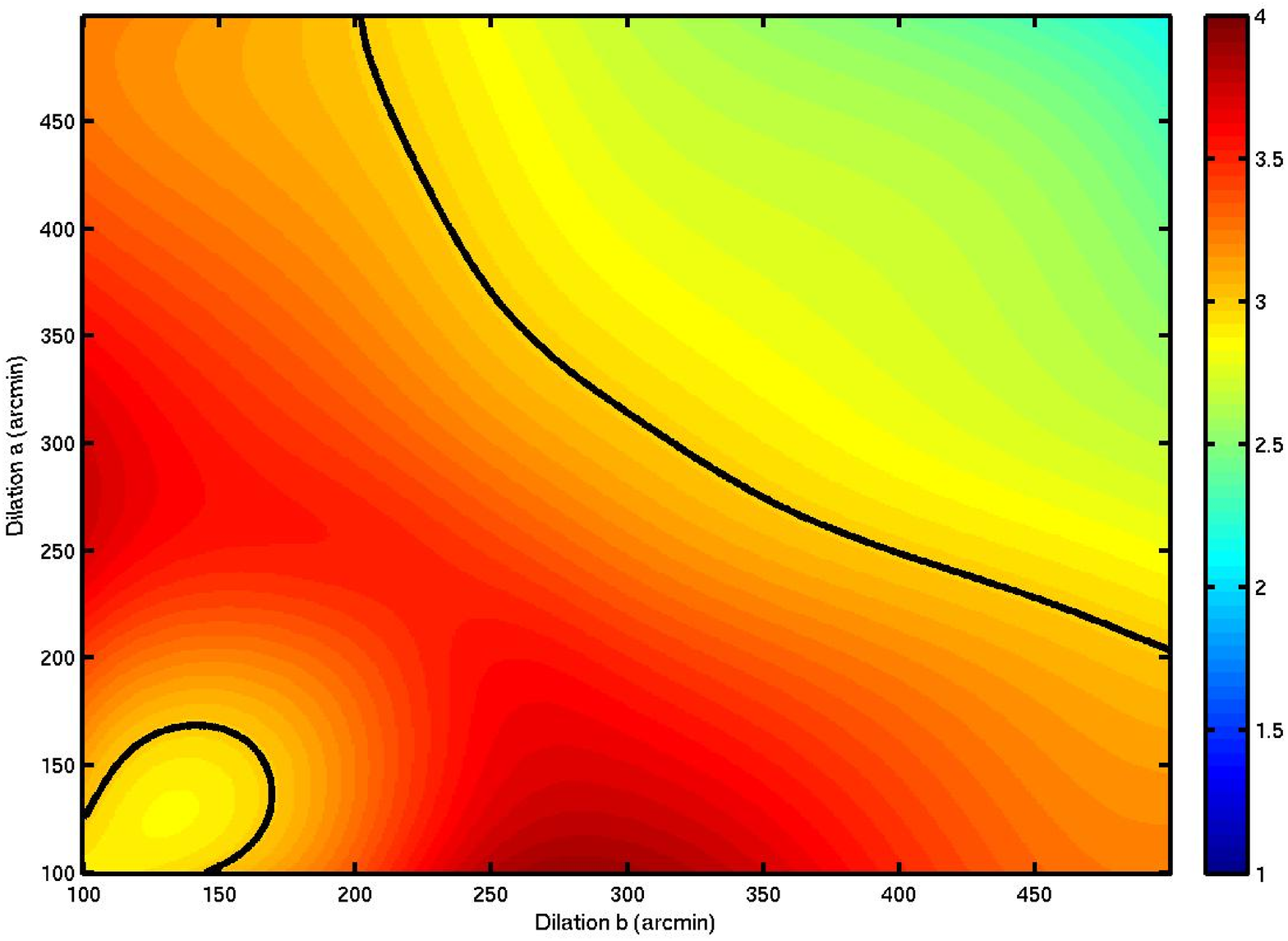}}
\subfigure[\sbw]
  {\includegraphics[clip=,width=\nsigmaplotwidth]{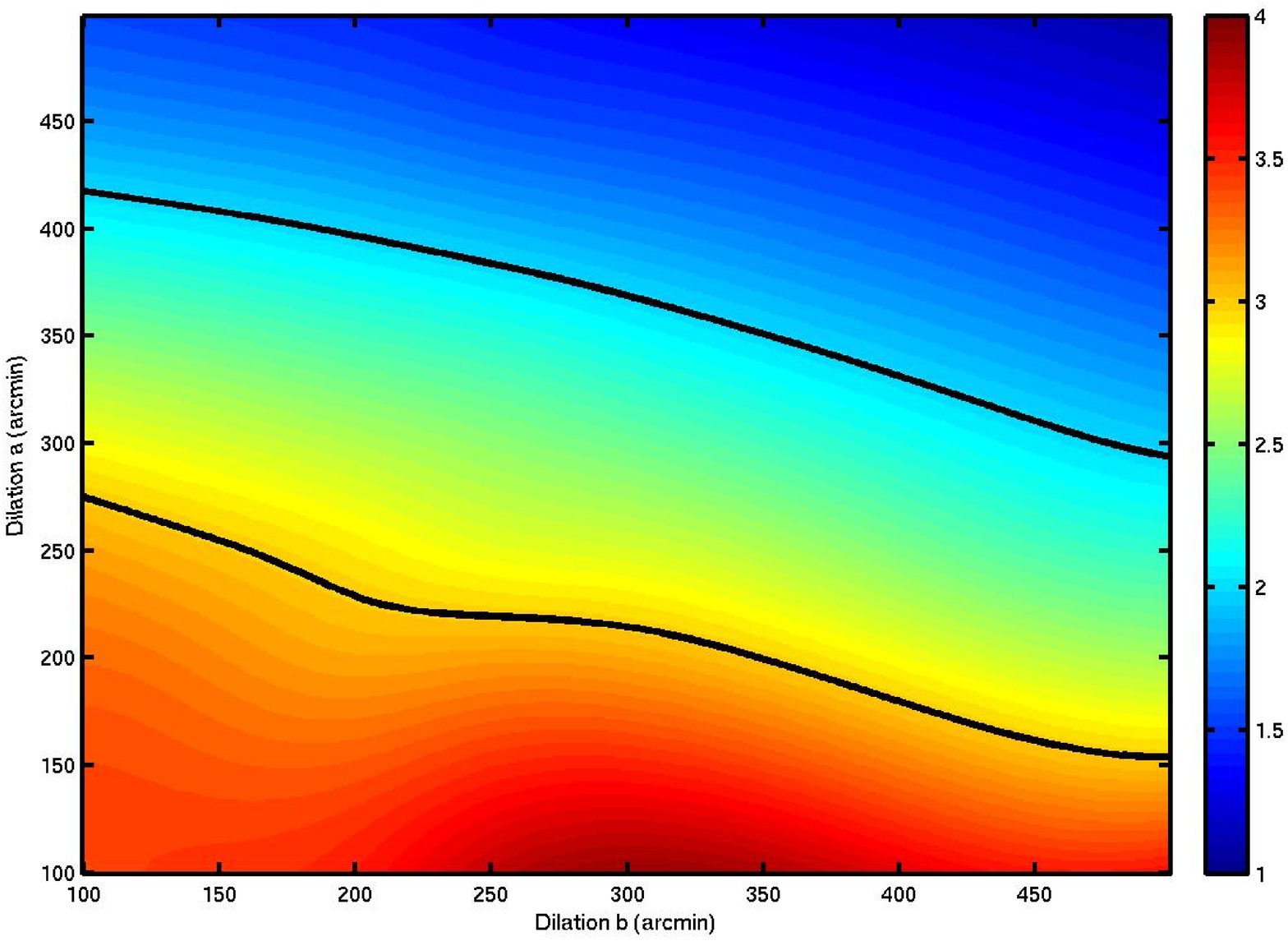}}
\caption{Wavelet covariance \nsigma\ surfaces.  Contours are shown for levels of two and three \nsigma.}
\label{fig:nsigma}
\end{figure}

\subsubsection{Foregrounds and systematics}

To test whether foregrounds or \wmap\ systematics are responsible for the correlation signal we examine the wavelet covariance obtained using both the separate \wmap\ bands and also combinations of difference maps constructed from the individual bands.
We have two Q-band maps at 40.7\ghz, two V-band maps at 60.8\ghz\ and four W-band maps at 93.5\ghz\ to construct test maps from (we subsequently use the notation Q1, Q2, \etc\ to denote the signal measured by each receiver and the single band letter to denote the sum of all maps measured for the given band, \eg\ Q=Q1+Q2).
Firstly, the wavelet covariance of the \nvss\ map with each of the Q-, V- and W-band maps is computed and is displayed in \fig{\ref{fig:bands}}.
For each case the wavelet covariance signal is essentially identical to the signal observed when using the co-added \wmap\ map (which is also plotted in \fig{\ref{fig:bands}} for comparison).
Secondly, we construct the difference maps \mbox{Q1-Q2}, V1-V2 and W1-W2+W3-W4 for each band from the signals measured by different receivers.  The \cmb\ signal and foreground contributions are completely removed from these noisy difference maps.  In addition, we examine the difference map W-V-Q which is free of CMB but has a clear foreground contribution.  The wavelet covariance of the \nvss\ map with each of these difference maps is computed and is displayed in \fig{\ref{fig:diff}}.  For all of these difference maps the correlation previously detected in the co-added \wmap\ is eliminated.

Any correlation between the \wmap\ and \nvss\ maps due to unremoved foreground emission in the \wmap\ data is expected to be frequency dependent, reflecting the emission law of the point source population (\eg\ \citealt{toffolatti:1998}).  However we observe identical wavelet covariance signals in each band (\fig{\ref{fig:bands}}) and do not detect any frequency dependence.  Furthermore, the W-V-Q difference map has minimal \cmb\ contribution but has a clear foreground contribution.  We do not detect any correlation using this map (\fig{\ref{fig:diff}}).  These findings suggest that the correlation signal that we detect in the co-added \wmap\ map is not due to unremoved foregrounds in the \wmap\ data (we show also in \sectn{\ref{sec:wav_loc}} that the detection is not due to a few localised regions, giving further evidence to support the claim that unremoved foregrounds are not responsible).

The wavelet covariance signal that we detect in the co-added \wmap\ map is present in all of the individual \wmap\ bands (\fig{\ref{fig:bands}}).  Furthermore, the covariance signal is eliminated in each of the individual band difference maps that contain no \cmb\ or foreground contributions (\fig{\ref{fig:diff}}).  Since all of the \wmap\ receivers produce an identical covariance signal that does not appear to be due to the noise artifacts of each receiver, we may conclude that it is unlikely that systematics are responsible for the detection made.

These preliminary tests show that it is unlikely that unremoved foreground emission or systematic effects in the \wmap\ data are responsible for the detection of a positive correlation between the \wmap\ and \nvss\ data.  It would appear that the detection is due to solely the \cmb\ contribution of the \wmap\ data.  This gives further support to the claim that the correlation we observe is produced by the \isw\ effect.

\setlength{\xcorrplotwidth}{60mm}

\begin{figure}
\centering
\subfigure[\smhw]
  {\includegraphics[clip=,width=\xcorrplotwidth]{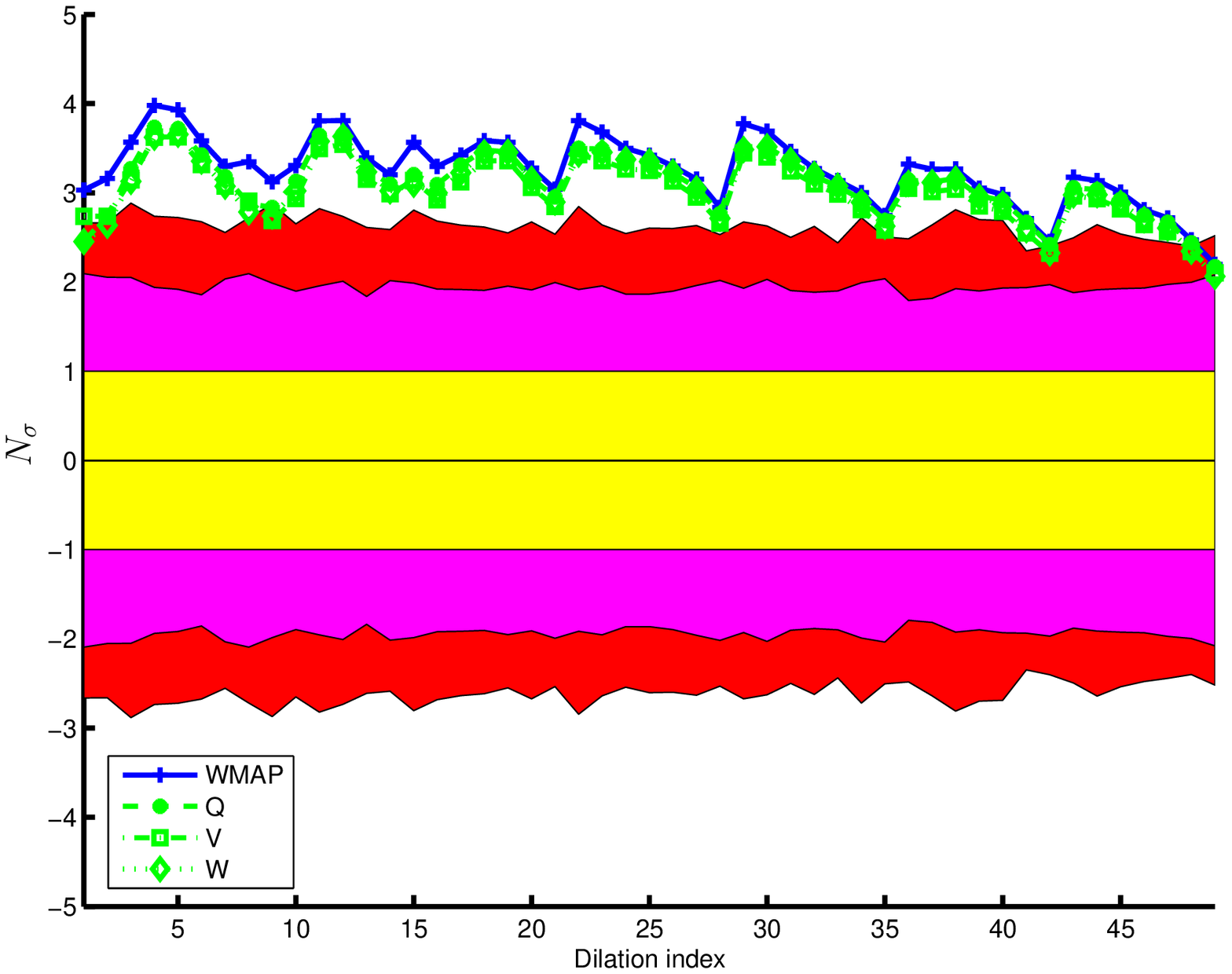}}
\subfigure[\sbw]
  {\includegraphics[clip=,width=\xcorrplotwidth]{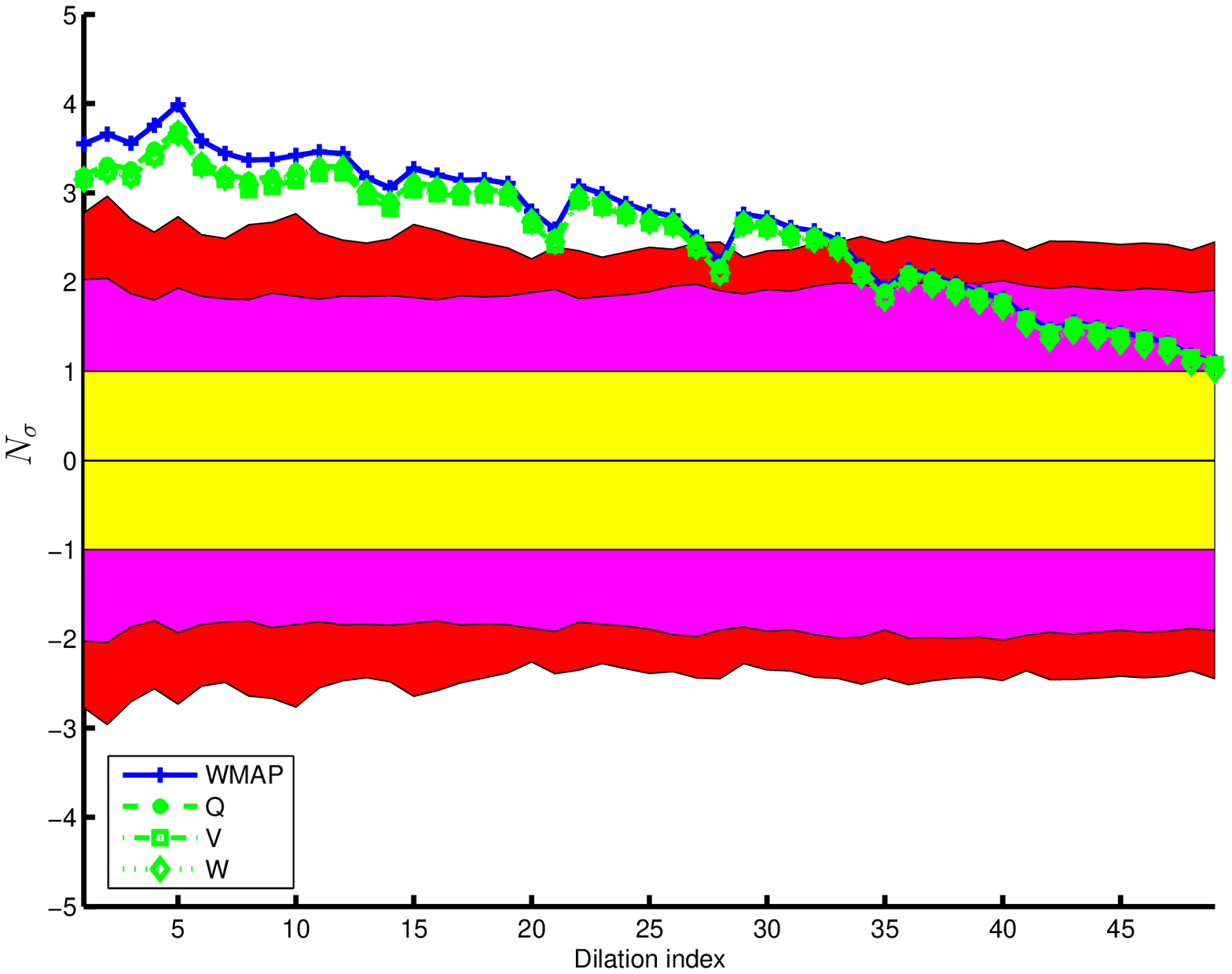}}
\caption{Wavelet covariance statistics in units of \nsigma\ computed for the \nvss\ map with individual \wmap\ band maps.}
\label{fig:bands}
\end{figure}

\begin{figure}
\centering
\subfigure[\smhw]
  {\includegraphics[clip=,width=\xcorrplotwidth]{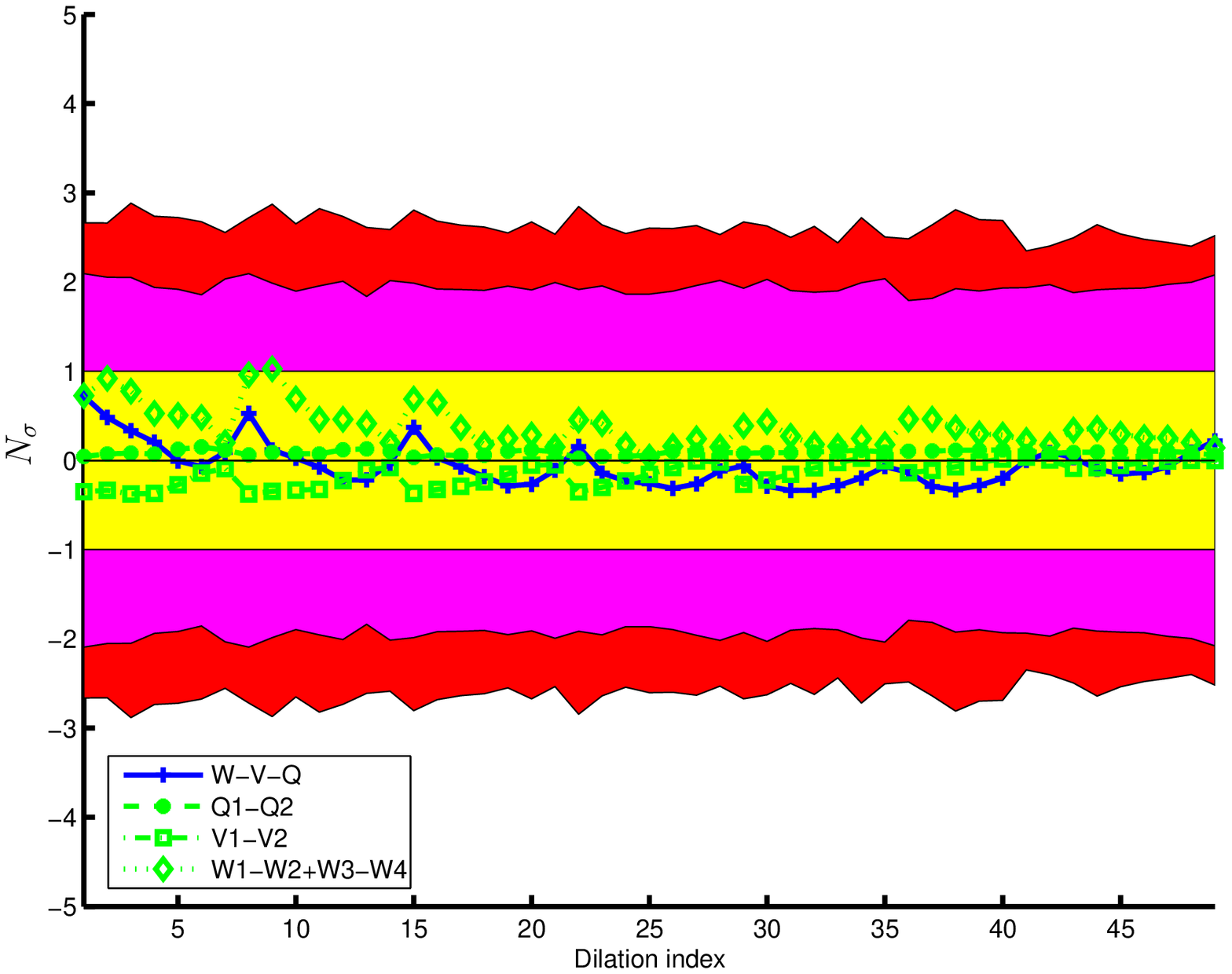}} 
\subfigure[\sbw]
  {\includegraphics[clip=,width=\xcorrplotwidth]{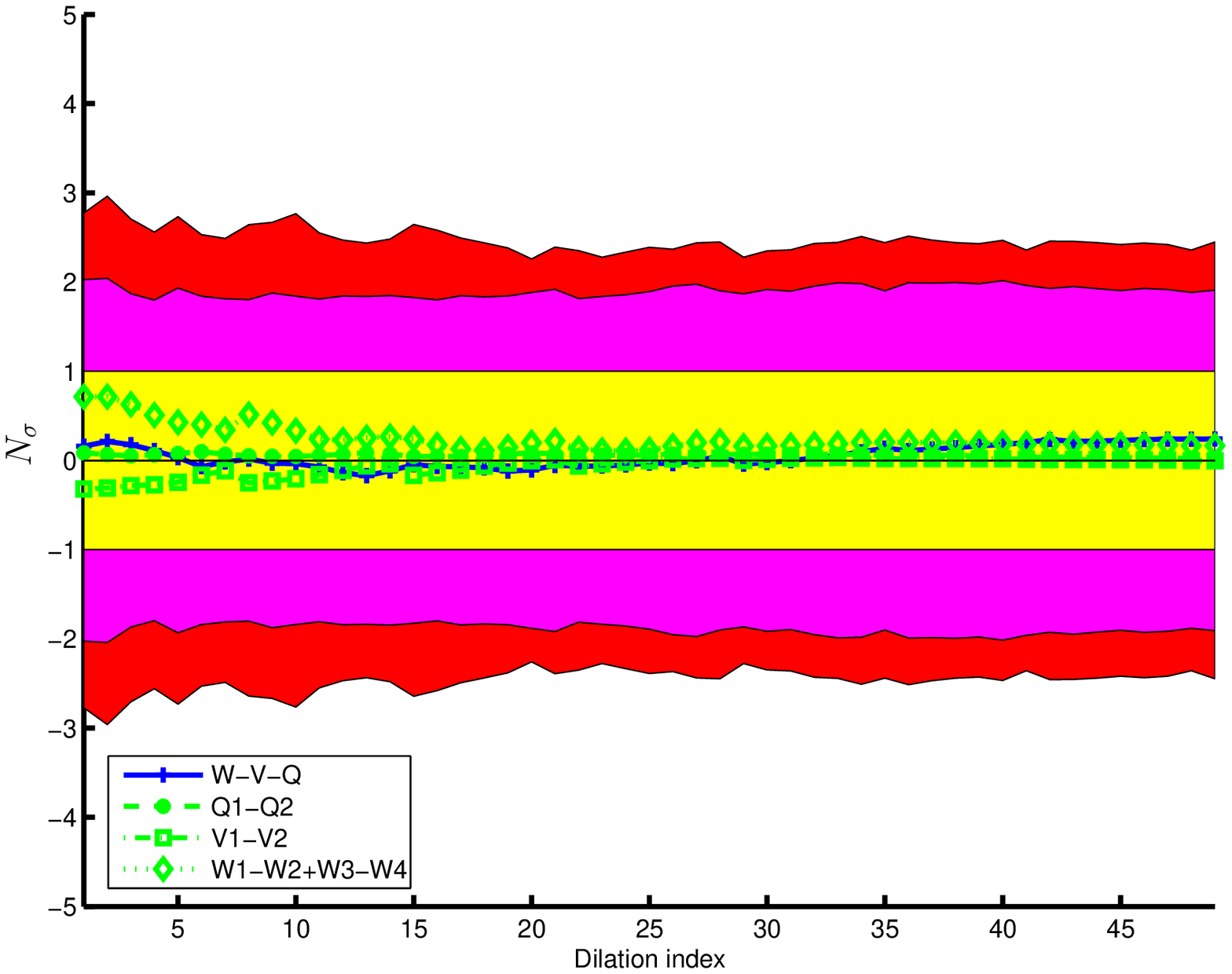}}
\caption{Wavelet covariance statistics in units of \nsigma\ computed for the \nvss\ map with \wmap\ band difference maps.}
\label{fig:diff}
\end{figure}

\subsubsection{Localised regions}
\label{sec:wav_loc}

A wavelet analysis inherently allows the spatial localisation of interesting signal characteristics.  The regions that contribute most strongly to the wavelet covariance detected may therefore be localised, not only in scale, but also in position and orientation on the sky.  

The wavelet covariance statistic is essentially the weighted mean of the wavelet domain product map constructed from the product of the \wmap\ and \nvss\ wavelet coefficients.  The regions that contribute most strongly to the covariance may thus be localised by thresholding the wavelet domain product map.  
In \fig{\ref{fig:xcorrprod}} we illustrate the localised regions detected once the wavelet product map is thresholded so that only those values that lie above $3\sigma$ remain.
For illustrational purposes we show for each wavelet only the localised maps corresponding to the scale and first orientation of the most significant covariance detection.  In the subsequent analysis all orientations are considered, thus \fig{\ref{fig:xcorrprod}} is illustrative rather than representing all of the localised regions that we detect.
Notice that the localised regions found are relatively evenly distributed over the sky (remember that we cannot search within the extended joint \wmap -\nvss\ mask, which becomes progressively larger on larger wavelet scales).  The localised maps found with the various wavelets also exhibit a fair degree of similarity.

A interesting next step is to determine whether these localised regions are the sole source of the wavelet covariance signal detected.  To examine this hypothesis we remove the localised regions detected on all scales and orientations (not just the ones corresponding to the maximum detections that are displayed in \fig{\ref{fig:xcorrprod}}) and repeat the analysis.  The wavelet covariance signals measured with and without the localised regions removed are shown in \fig{\ref{fig:xcorrloc}}.  Once the localised regions are removed the covariance signal is reduced in significance, as one would expect since those regions that contribute most strongly are removed, but the covariance signal is not eliminated.  Indeed, the signal remains on certain scales at almost the 95\% significance level.  We conclude that the localised regions that we detect are \emph{not} the sole source of correlation between the \wmap\ and \nvss\ data.  Our findings are consistent intuitively with the predicitions of the \isw\ effect, namely that we would expect any observed correlation to be due to weak correlations over the entire sky rather than a few localised regions.  Furthermore, the 
aggregate nature of the signal detected provides further evidence to suggest that foreground contamination in the data is not responsible.
\citet{boughn:2005} make similar conclusions that the correlation signal they detect is due to aggregate correlations over the entire sky rather than a few localised regions.

As a final test of the integrity of the data we examine in more detail a few of the most significant localised regions that we detect.  We examine 18 of the localised regions illustrated in \fig{\ref{fig:xcorrprod}}, selected from the regions that contain the greatest wavelet coefficient product map values, 
in both the \nvss\ and Bonn 1420MHz \citep{reich:1982,reich:1986} radio surveys (the latter data set affords better visualisation of the \lss\ and enables one to highlight regions in the \nvss\ data for closer examination).  The regions selected are defined in \tbl{\ref{tbl:regions}} and \fig{\ref{fig:regions}}.
We do not notice any difference in the \nvss\ data between our localised regions and regions selected at random.  The localised regions we detect do not, in general, correspond to regions with particularly bright point sources that could have caused contamination of the data.

All of the tests that we have performed indicate that the correlation detected is indeed due to the \isw\ effect and not due to systematics or foreground contamination.  Next we use our positive detection of the \isw\ effect to place constraints on dark energy.


\setlength{\mapplotwidth}{60mm}

\begin{figure}
\centering
\subfigure[Symmetric \smhw ; dilation $(\scaleab)=(250\arcmin,250\arcmin)$]
  {\includegraphics[bb= 0 40 800 440,clip=,width=\mapplotwidth]{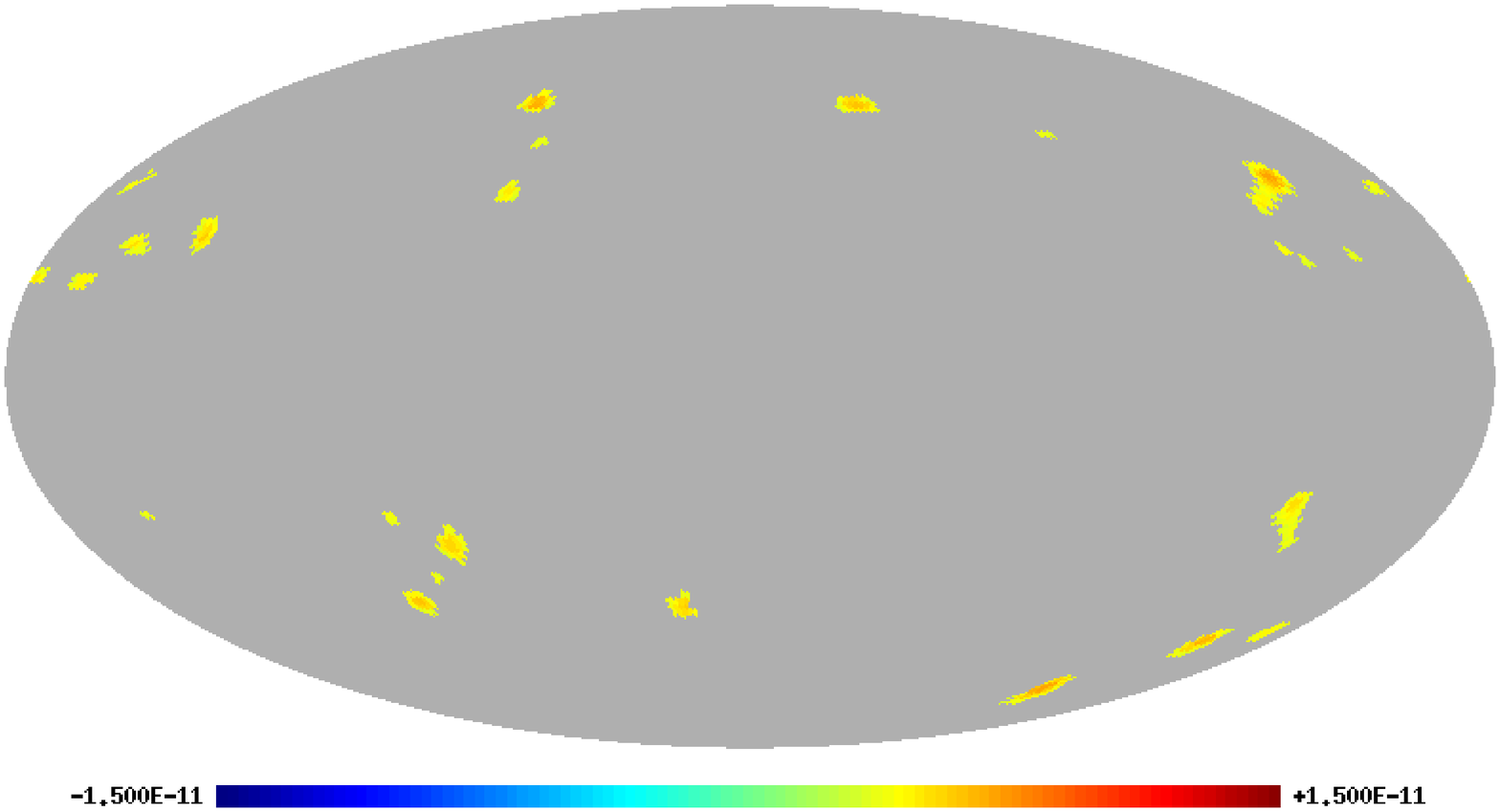}}
\subfigure[Elliptical \smhw ; dilation $(\scaleab)=(100\arcmin,300\arcmin)$]{\includegraphics[bb= 0 40 800 440,clip=,width=\mapplotwidth]{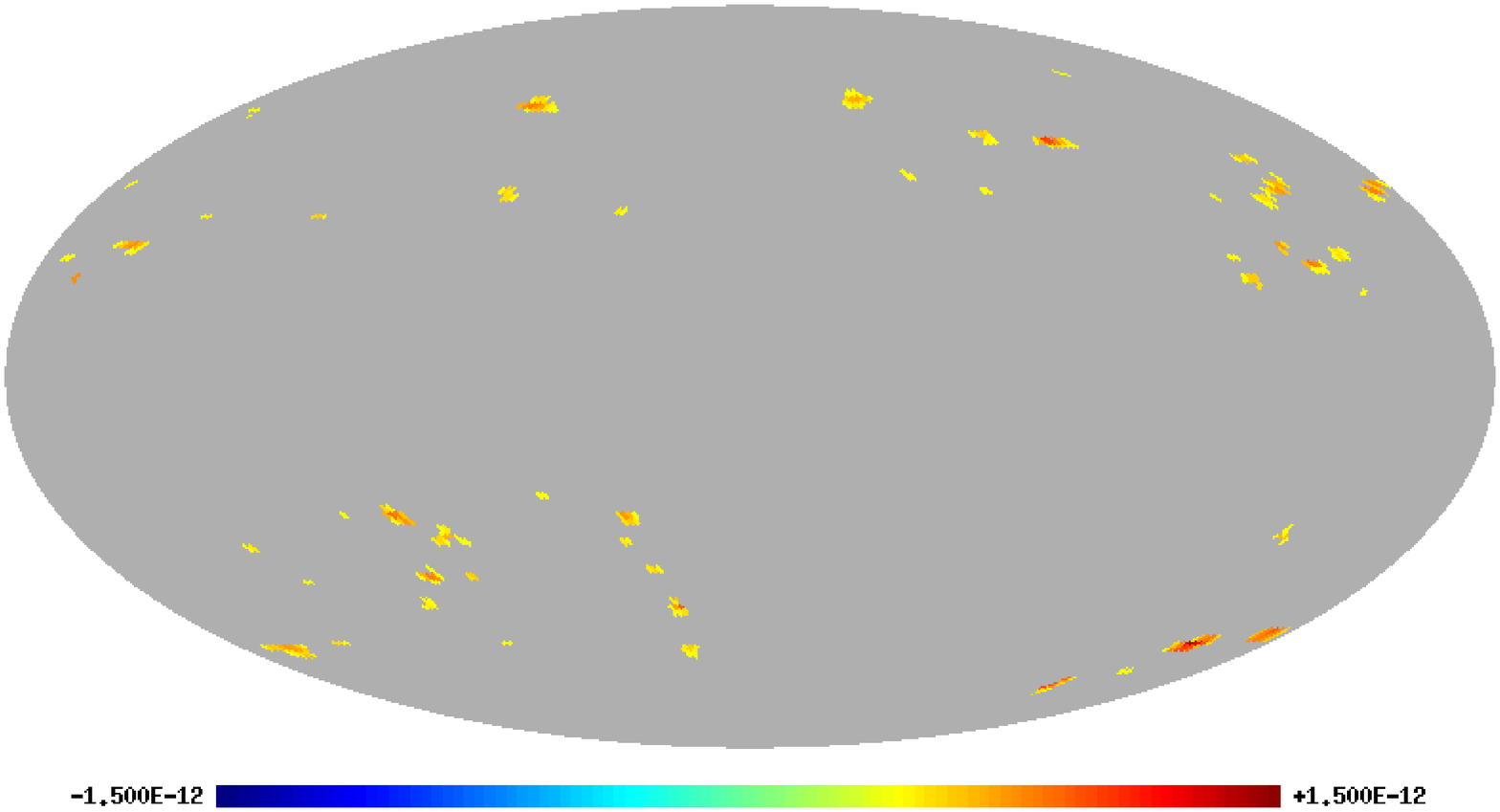}}
\subfigure[\sbw ; dilation $(\scaleab)=(100\arcmin,300\arcmin)$]{\includegraphics[bb= 0 40 800 440,clip=,width=\mapplotwidth]{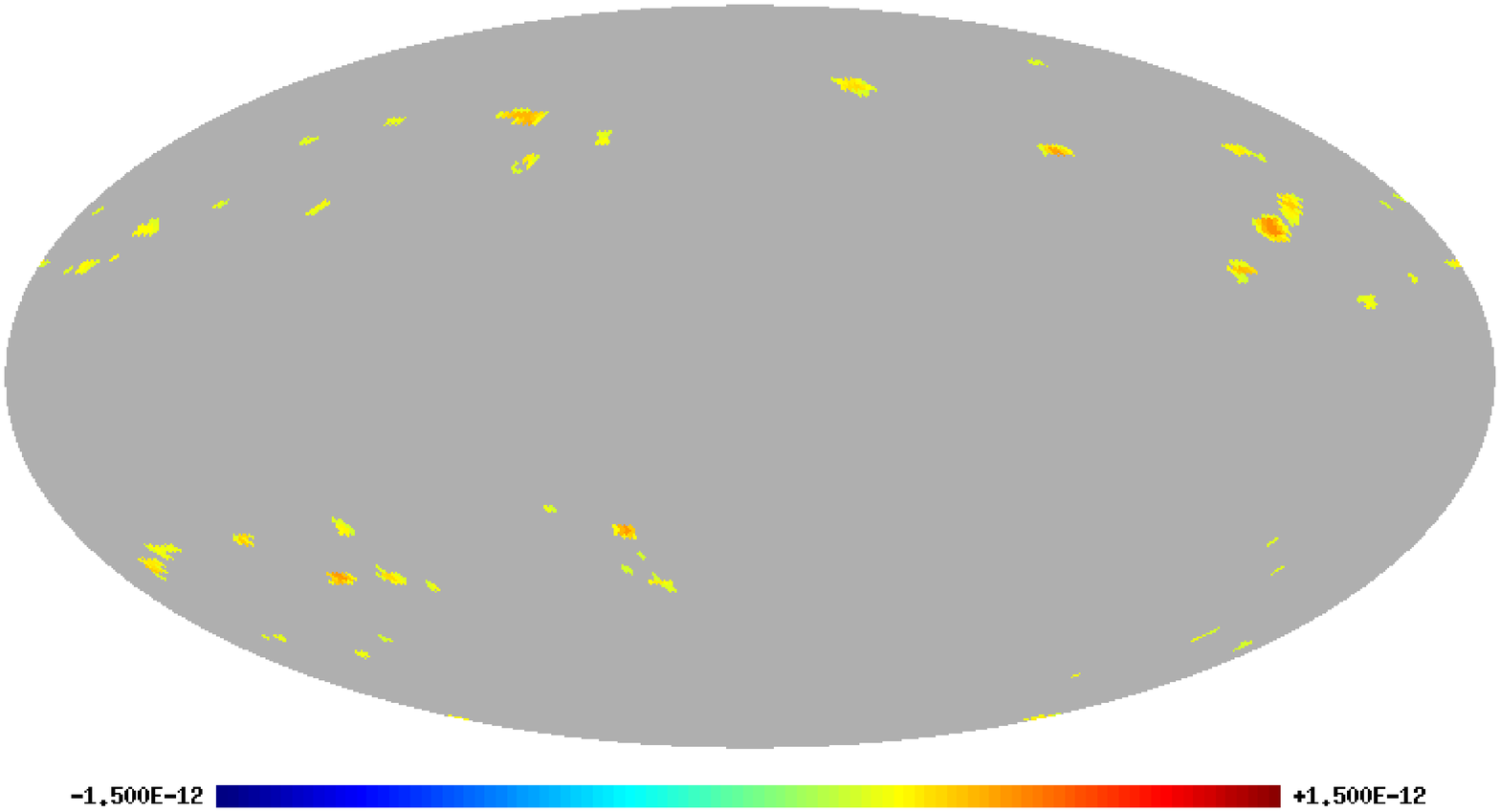}}
\caption{Localised \wmap -\nvss\ wavelet coefficient product maps thresholded at $3\sigma$.  The regions remaining show those areas that contribute most strongly to the positive wavelet covariance signal detected.
Localised product maps are only shown for the first orientation for the scale of the most significant correlation detection made with each wavelet.}
\label{fig:xcorrprod}
\end{figure}

\setlength{\xcorrplotwidth}{55mm}

\begin{figure*}
\begin{minipage}{\textwidth}
\centering
\mbox{
\subfigure[Symmetric \smhw]
  {\includegraphics[clip=,width=\xcorrplotwidth]{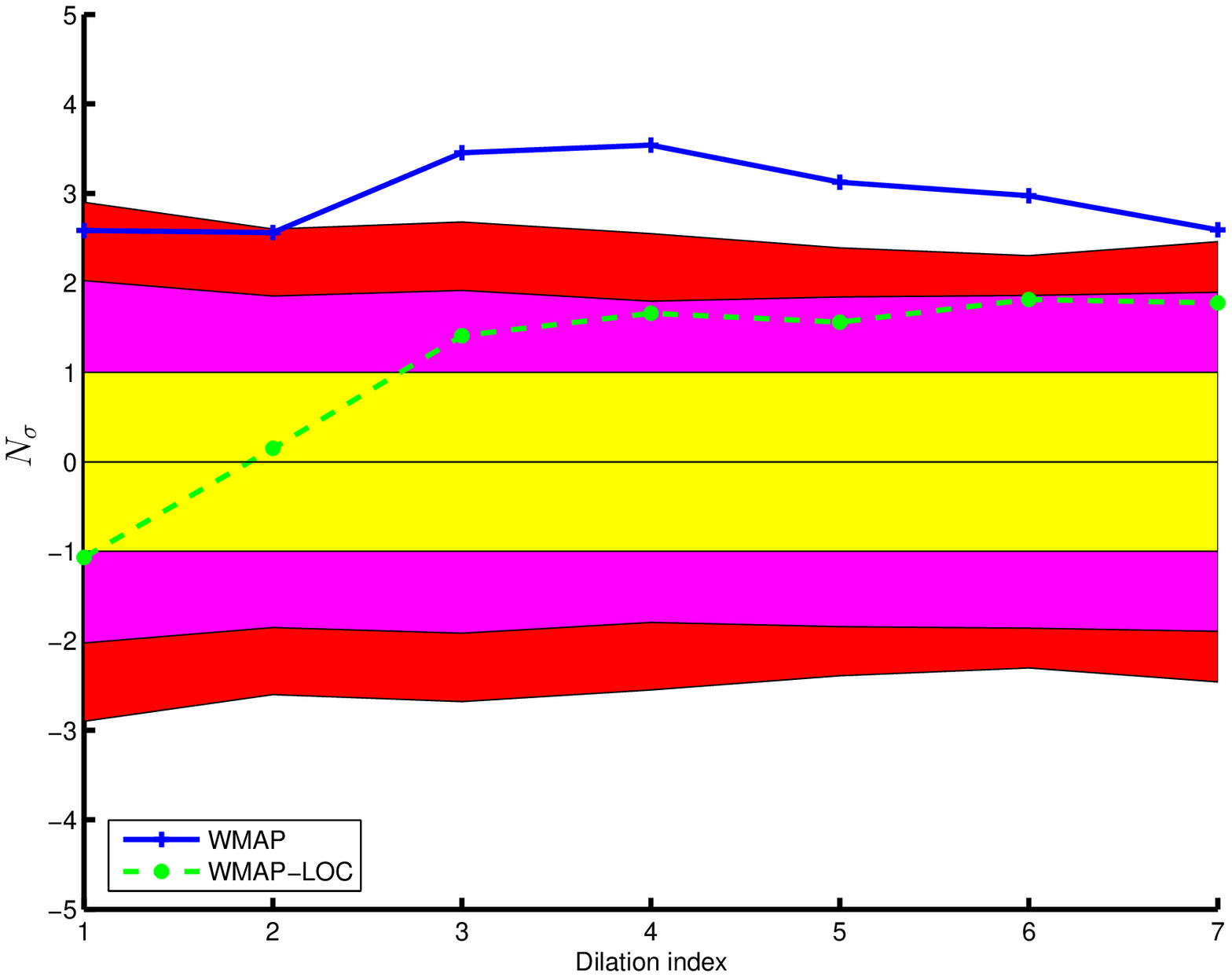}} \quad
\subfigure[Elliptical \smhw]
  {\includegraphics[clip=,width=\xcorrplotwidth]{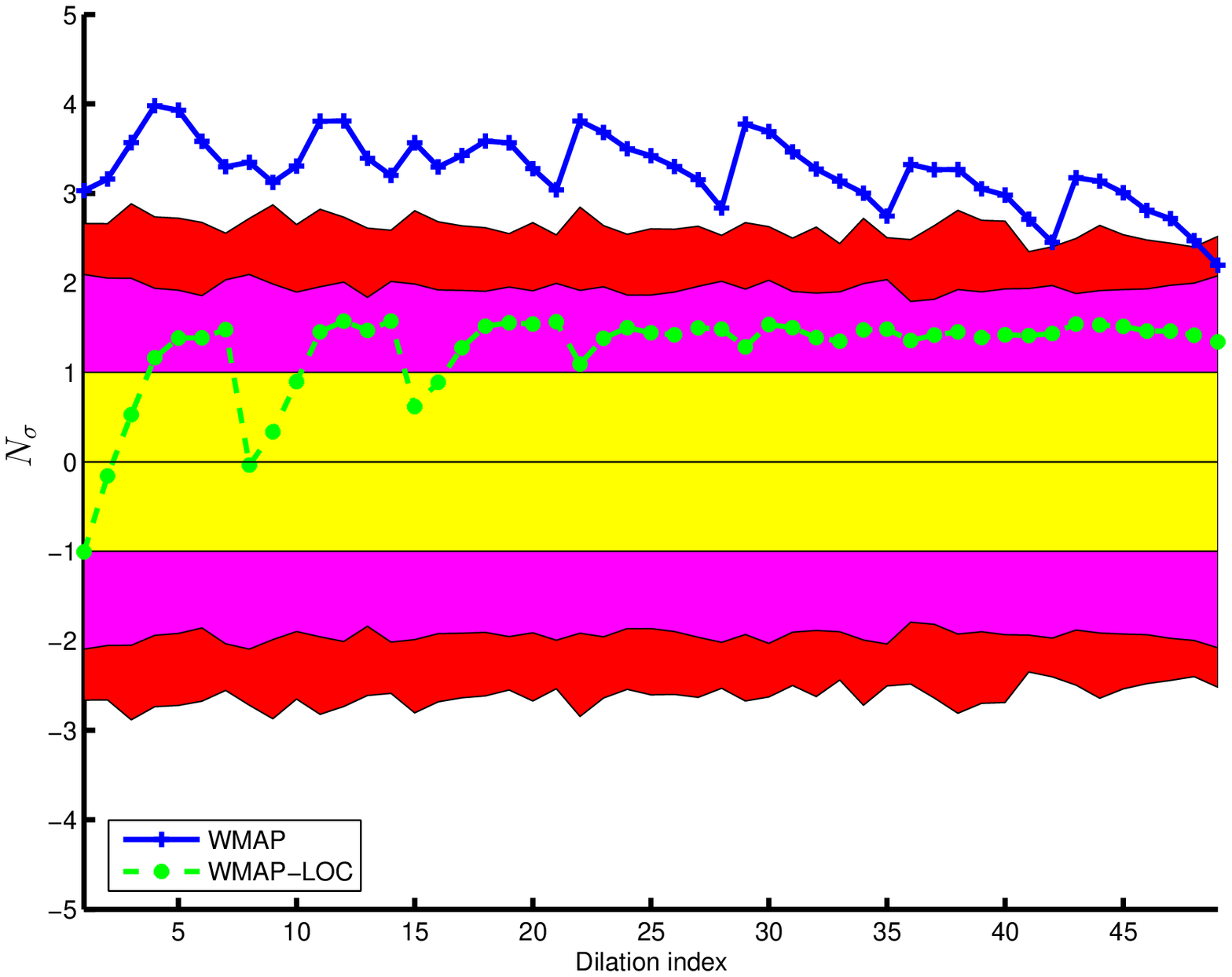}} \quad
\subfigure[\sbw]
  {\includegraphics[clip=,width=\xcorrplotwidth]{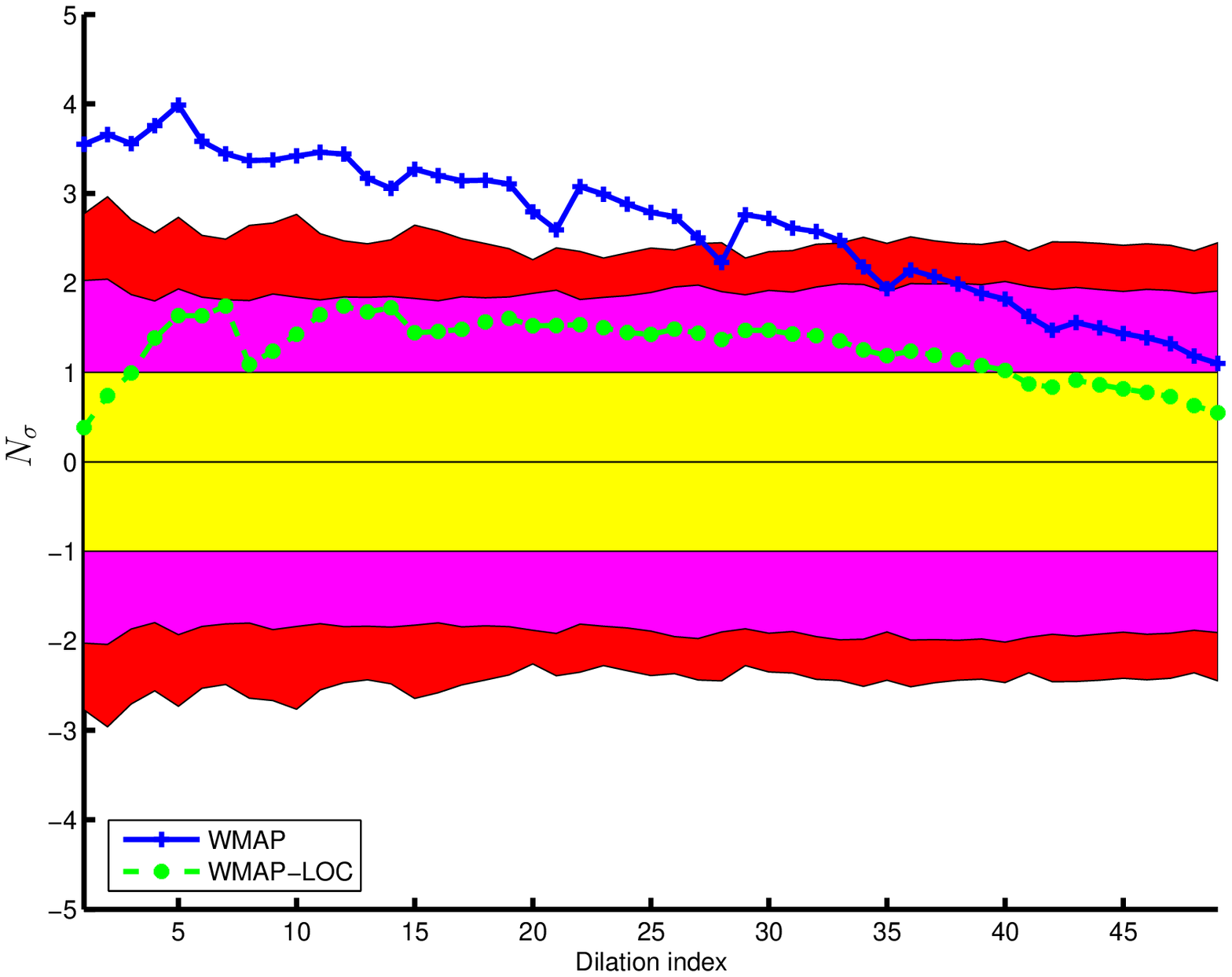}}
}
\caption{Wavelet covariance statistics computed for the \wmap\ and \nvss\ data with (green/light-grey dots) and without (blue/dark-grey pluses) the localised regions that contribute most strongly to the covariance removed.  The covariance signal is reduced in significance when the localised regions are removed, as one would expect, but it is certainly not eliminated suggesting that the localised regions are \emph{not} the sole source of the wavelet covariance.}
\label{fig:xcorrloc}
\end{minipage}
\end{figure*}

\begin{figure}
\centering
\includegraphics[clip=,width=\mapplotwidth]{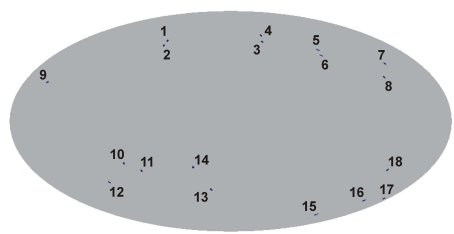}
\caption{Approximate localised regions flagged for closer examination (see \tbl{\ref{tbl:regions}} for more details).}
\label{fig:regions}
\end{figure}

\begin{table}
\caption[Localised regions flagged by the \isw\ analysis for closer examination]{Approximate localised regions flagged for closer examination (see \fig{\ref{fig:regions}} for a map of these regions).  The wavelets 
that flag each region are listed also (the same regions are flagged by both the symmetric and elliptical \smhw).  Although these regions are specified precisely they represent only the approximate position of the regions localised by the wavelet analysis, which is often of the order of a number of degrees.}
\label{tbl:regions}
\centering
\footnotesize
\begin{tabular}{ccccc} \hline
Region & \multicolumn{2}{c}{Location} & \multicolumn{2}{c}{Flagged by wavelet} \\
& Longitude & Latitude & \smhw & \sbw \\ \hline
 1 &  75\degrees &  57\degrees & $\checkmark$ & $\times$ \\
 2 &  75\degrees &  53\degrees & $\times$ & $\checkmark$ \\
 3 & 323\degrees &  56\degrees & $\checkmark$ & $\times$ \\
 4 & 321\degrees &  62\degrees & $\times$ & $\checkmark$ \\
 5 & 267\degrees &  50\degrees & $\checkmark$ & $\times$ \\
 6 & 268\degrees &  45\degrees & $\times$ & $\checkmark$ \\
 7 & 213\degrees &  40\degrees & $\checkmark$ & $\times$ \\
 8 & 223\degrees &  30\degrees & $\times$ & $\checkmark$ \\
 9 & 160\degrees &  26\degrees & $\checkmark$ & $\checkmark$ \\
10 &  94\degrees & -28\degrees & $\checkmark$ & $\times$ \\
11 &  81\degrees & -34\degrees & $\checkmark$ & $\times$ \\
12 & 118\degrees & -42\degrees & $\times$ & $\checkmark$ \\
13 &  20\degrees & -48\degrees & $\checkmark$ & $\times$ \\
14 &  34\degrees & -31\degrees & $\times$ & $\checkmark$ \\
15 & 230\degrees & -68\degrees & $\checkmark$ & $\times$ \\
16 & 204\degrees & -56\degrees & $\checkmark$ & $\checkmark$ \\
17 & 186\degrees & -54\degrees & $\checkmark$ & $\checkmark$ \\
18 & 218\degrees & -33\degrees & $\checkmark$ & $\times$ \\
\hline
\end{tabular}
\end{table}

\subsection{Constraints on $\olambda$ and $\w$}
\label{sectn:constraintsboth}

In a flat universe the \isw\ effect is present only in the presence of dark energy.  We may therefore use our detection of the \isw\ effect to constrain dark energy parameters by comparing the theoretically predicted wavelet covariance signal for different cosmological models with that measured from the data.  In particular, we constrain the vacuum energy density \olambda\ and the equation-of-state parameter \w\ (assuming the equation-of-state parameter does not evolve with redshift within the epoch of interest).  We probe the parameter ranges $0<\olambda<0.95$ and $-2<\w<0$.
For the other cosmological parameters we use the concordance model values \citep{spergel:2003}.  We use a bias parameter value of $b=1.6$.  This is the best choice consistent with the concordance model that is within the range favoured by the \nvss\ data \citep{boughn:2002,boughn:2004} when adopting the RLF1 model proposed by \citet{dunlop:1990}.  In any case, \citet{vielva:2005} find dark energy constraints to be insensitive to changes in the bias within the range $1.4 < b < 1.8$.  We also use the \citet{dunlop:1990} RLF1 model to describe the \dndz\ distribution since it fits the \nvss\ galaxy autocorrelation function well, as previously shown by \citet{boughn:2002,boughn:2004} and \citet{nolta:2004}.

To compare the wavelet covariance of the data with the theoretical predictions made by different cosmological models we compute the $\chisqd$ for parameters $\baypar = (\olambda,\w)$, defined by
\begin{equation}
\chisqd(\baypar) = 
\left[ \wcovvect^{\ndlab\tplab}(\baypar) - \wcovestvect^{\ndlab\tplab} \right]^{\rm T}
{\rm C}^{-1}
\left[ \wcovvect^{\ndlab\tplab}(\baypar) - \wcovestvect^{\ndlab\tplab} \right]
\spcend ,
\end{equation}
where $\wcovvect^{\ndlab\tplab}(\baypar)$ is the predicted covariance vector for the given cosmological model and 
$\wcovestvect^{\ndlab\tplab}$ is the wavelet covariance vector of the data.  These vectors are constructed by concatenating the wavelet covariance values computed for the considered dilations and orientations into vectors.  The matrix ${\rm C}$ is a similarly ordered covariance matrix of the wavelet covariance statistics, computed from the 1000 Monte Carlo simulations.  Making a Gaussian approximation for the likelihood, we may write the likelihood function as $\lhood(\baypar) \propto \exp \left[ - \chisqd(\baypar)/2 \right]$.  The likelihood surfaces for the parameter ranges discussed for both of the \smhw\ and the \sbw\ are illustrated in \fig{\ref{fig:pdfall}}.  Marginalised likelihood distributions are also shown.
Parameter estimates are made from the mean of the marginalised distributions, with 68\%, 95\% and 99\% confidence regions constructed to ensure the required probability is met in the tails of the distribution.  The parameter estimates and errors made with the various wavelets are as follows:
$\olambda=0.63_{-0.17}^{+0.18}$, $\w=-0.77_{-0.36}^{+0.35}$ using the \smhw ;
$\olambda=0.52_{-0.20}^{+0.20}$, $\w=-0.73_{-0.46}^{+0.42}$ using the \sbw.
Within error bounds these parameter estimates are consistent with one another and also with estimates made from numerous other analysis procedures and data sets.  The errors we obtain on the parameter estimates are relatively large.  Although wavelets perform very well when attempting to detect the \isw\ effect since one may probe different scales, positions and orientations, once all information is incorporated to compute the likelihood surface the performance of a wavelet analysis is comparable to other linear techniques, as expected\footnote{The results of different linear analyses, however, will not be exactly identical due to differences in masking.} (see \citet{vielva:2005} for a detailed discussion and comparison of real, harmonic and (azimuthally symmetric) wavelet space techniques for detecting correlations due to the \isw\ effect).


\newlength{\pdfwidth}
\setlength{\pdfwidth}{70mm}

\begin{figure*}
\begin{minipage}{\textwidth}
\centering
  \mbox{
  \subfigure[Full likelihood surface obtained using  \smhw]{\includegraphics[width=\pdfwidth]{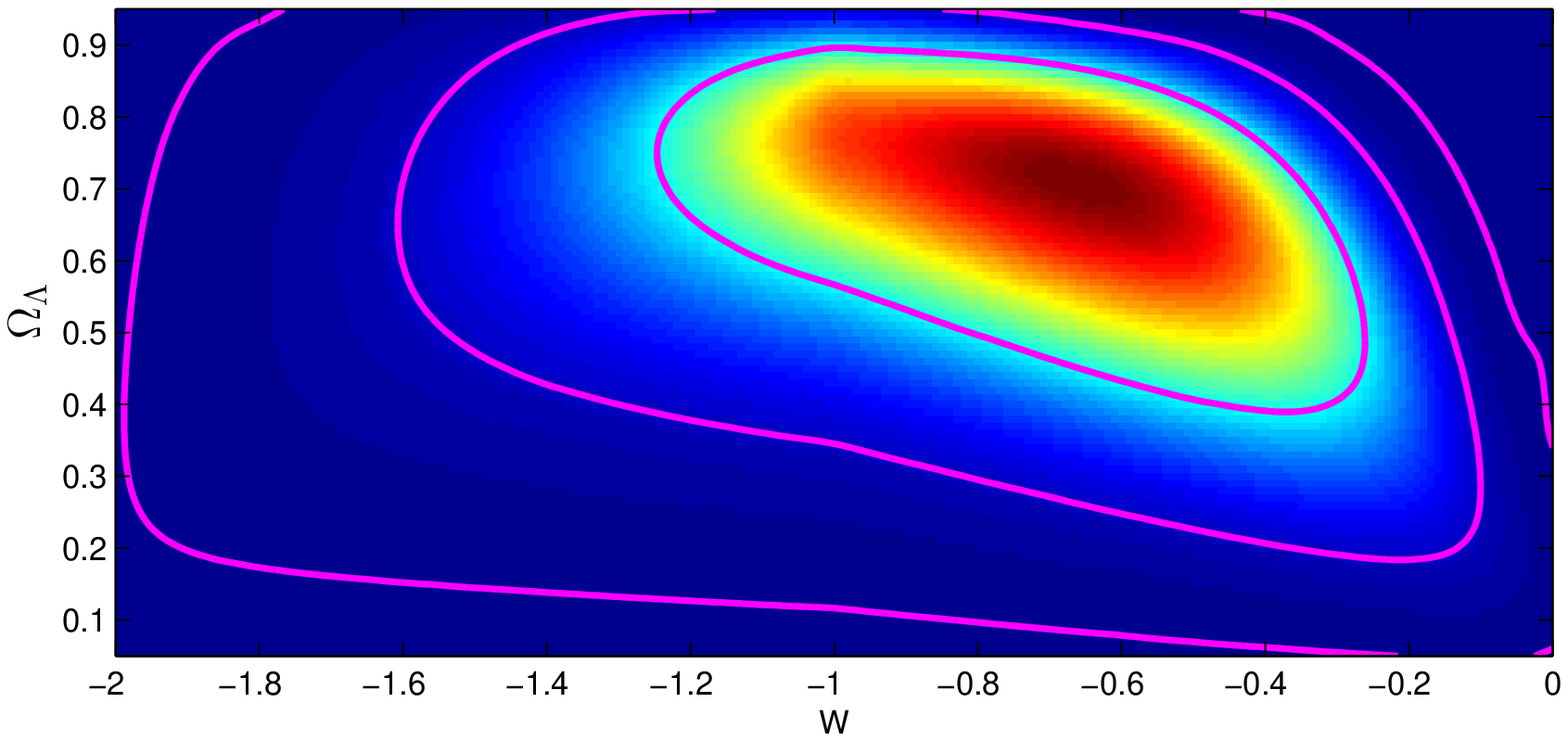}} \quad
  \subfigure[Full likelihood surface obtained using \sbw]{\includegraphics[width=\pdfwidth]{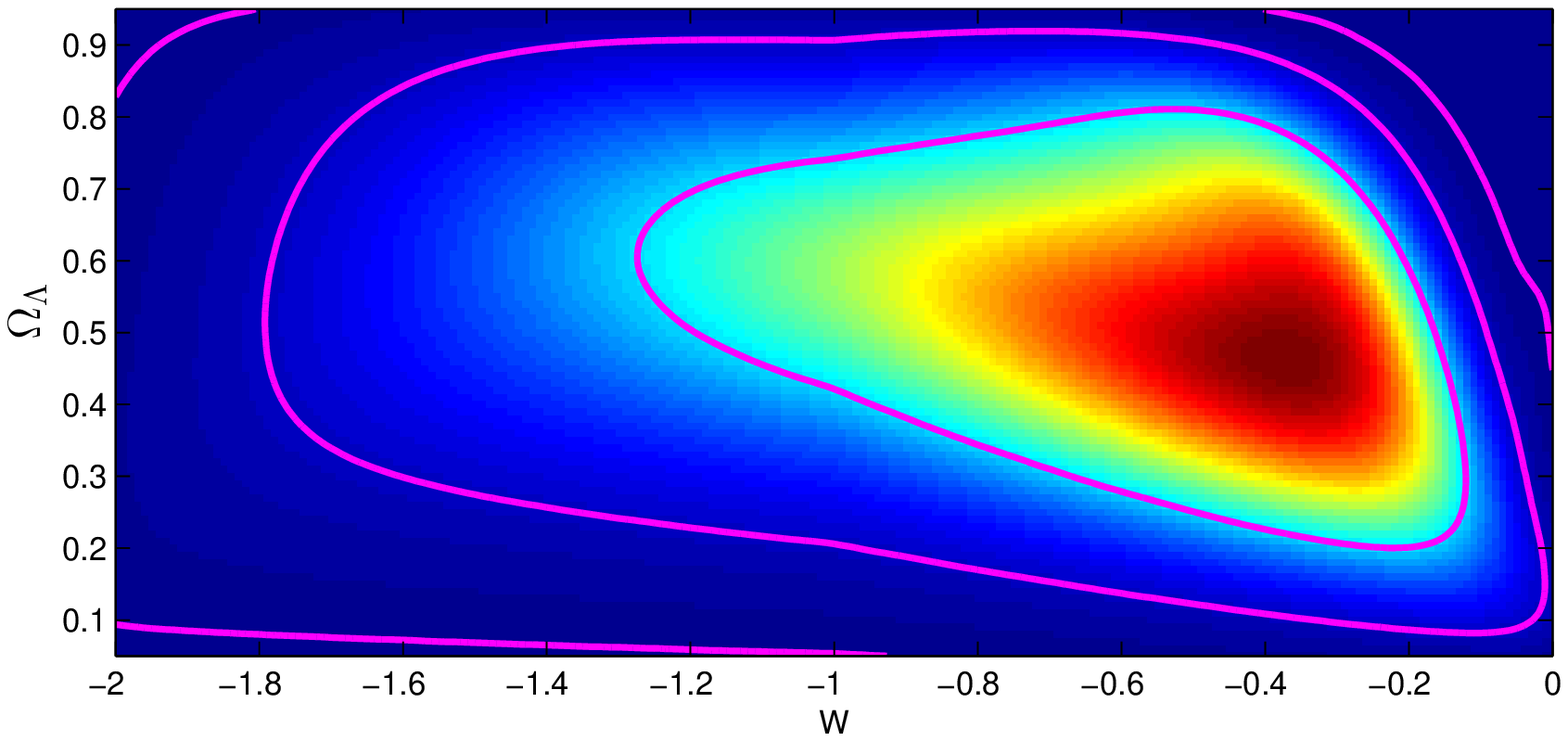}}
}
  \mbox{
  \subfigure[Marginalised distribution for \olambda\ obtained using \smhw]{\includegraphics[width=\pdfwidth]{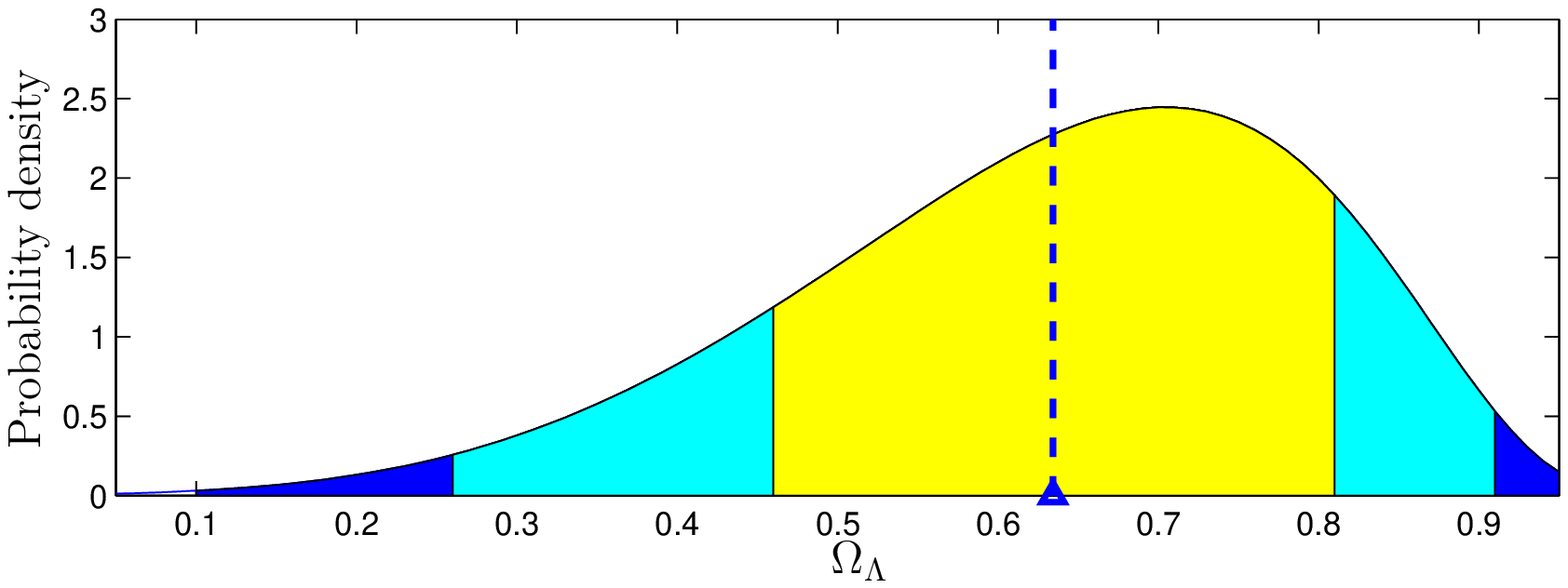}} \quad
  \subfigure[Marginalised distribution for \olambda\ obtained using \sbw]{\includegraphics[width=\pdfwidth]{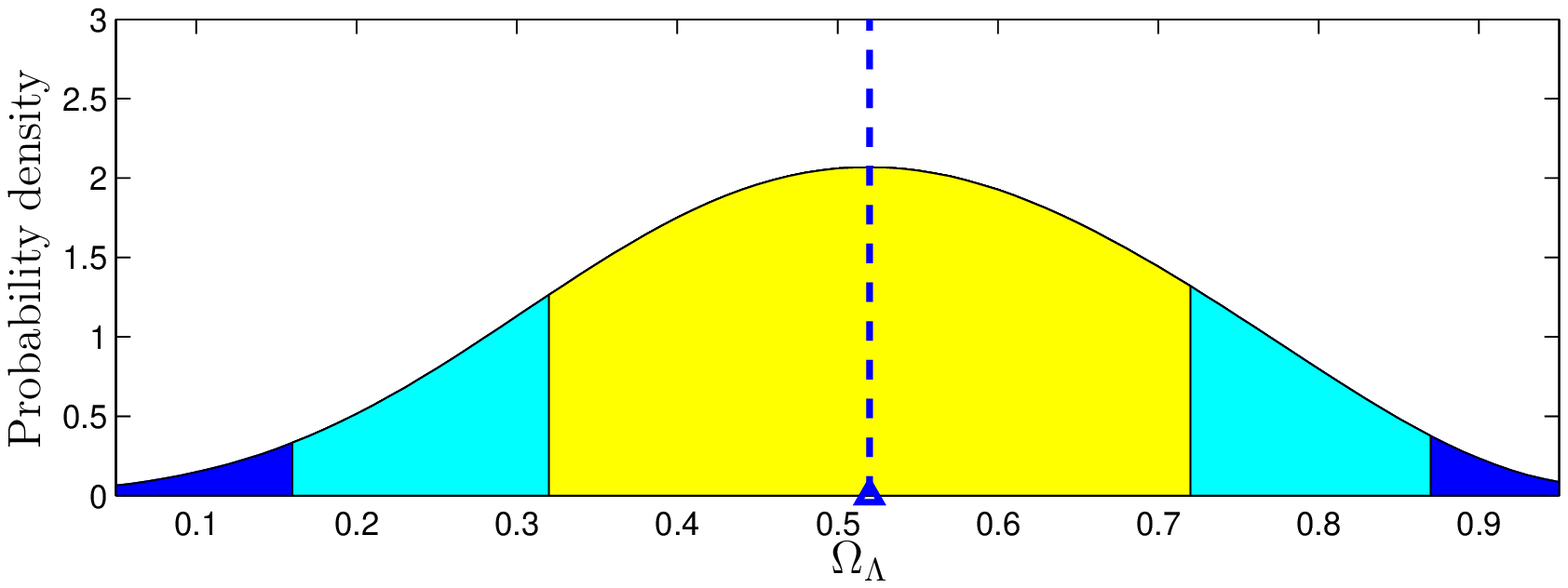}}
  }
  \mbox{
  \subfigure[Marginalised distribution for \w\ obtained using \smhw]{\includegraphics[width=\pdfwidth]{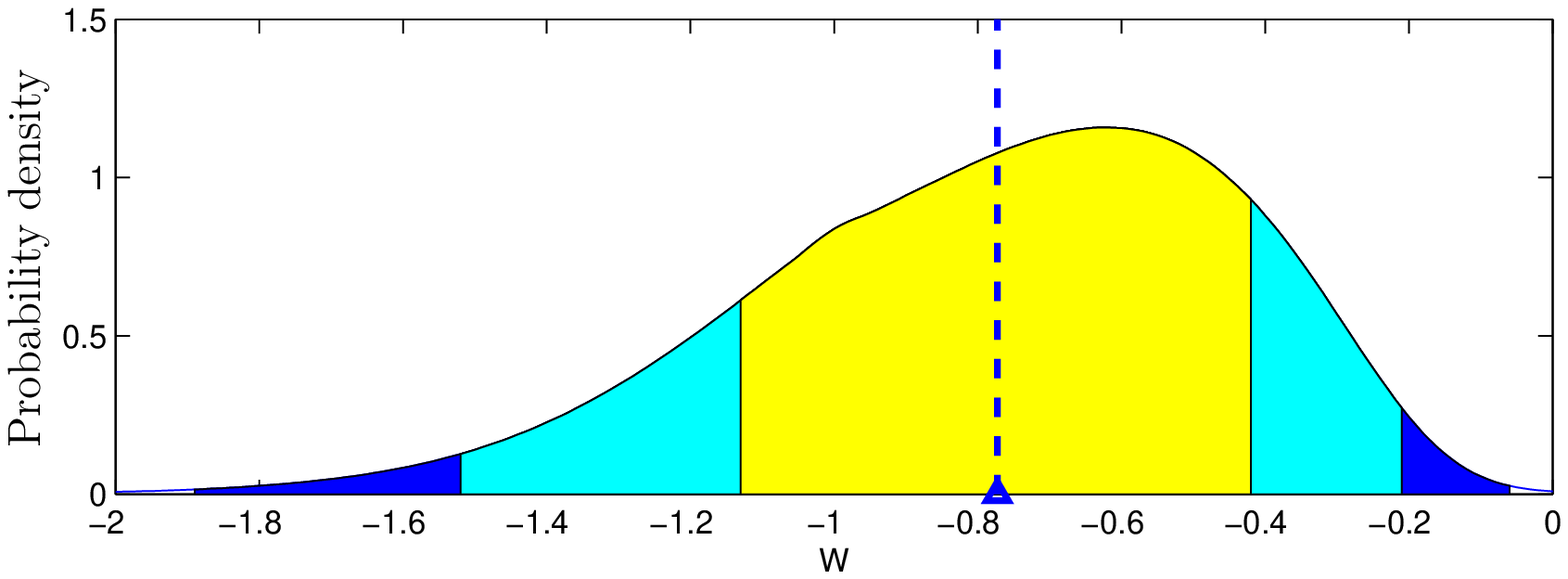}} \quad
  \subfigure[Marginalised distribution for \w\ obtained using \sbw]{\includegraphics[width=\pdfwidth]{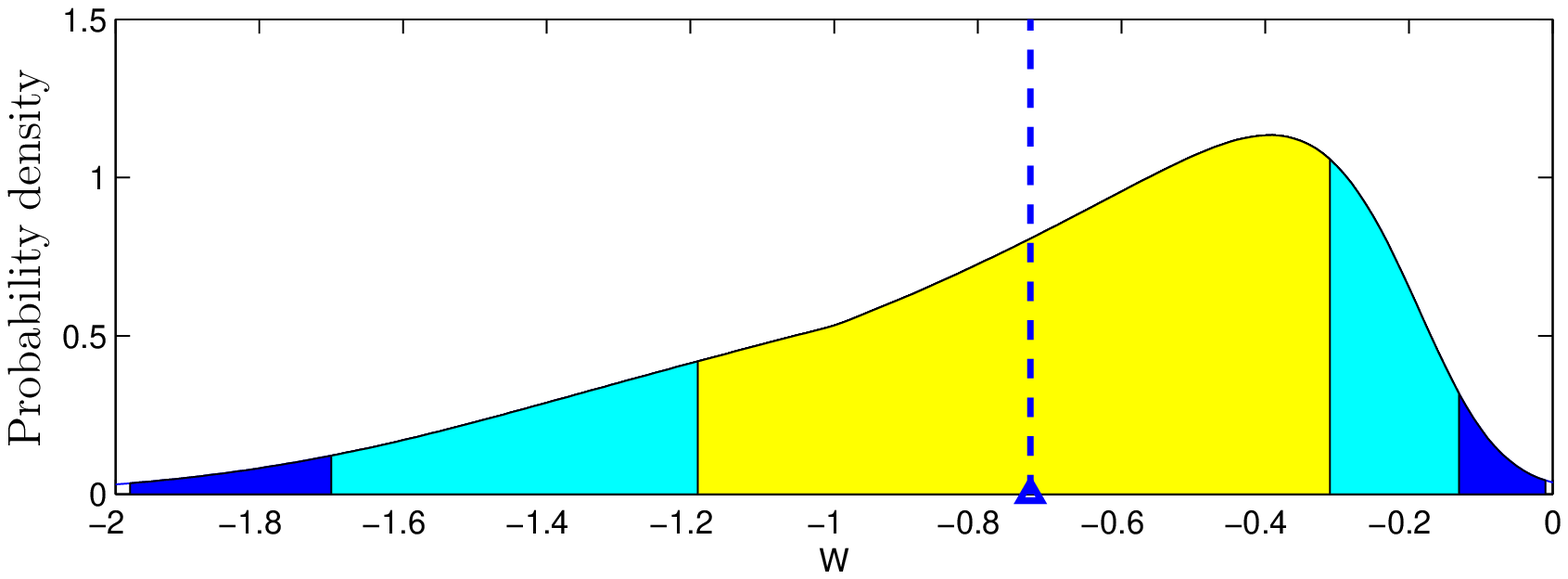}}
  }
\caption{Likelihood surfaces for parameters $\baypar = (\olambda,\w)$.  The full likelihood surfaces are shown in the top row of panels, with 68\%, 95\% and 99\% confidence contours also shown.  Marginalised distributions for each parameter are shown in the remaining panels, with 68\% (yellow/light-grey), 95\% (light-blue/grey) and 99\% (dark-blue/dark-grey) confidence regions also shown.  The parameter estimates made from the mean of the marginalised distribution is shown by the triangle and dashed line.}
\label{fig:pdfall}
\end{minipage}
\end{figure*}

\subsection{Constraints on $\olambda$}

\begin{figure}
\centering
\subfigure[Likelihood distribution for \olambda\ obtained using \smhw]{\includegraphics[width=\pdfwidth]{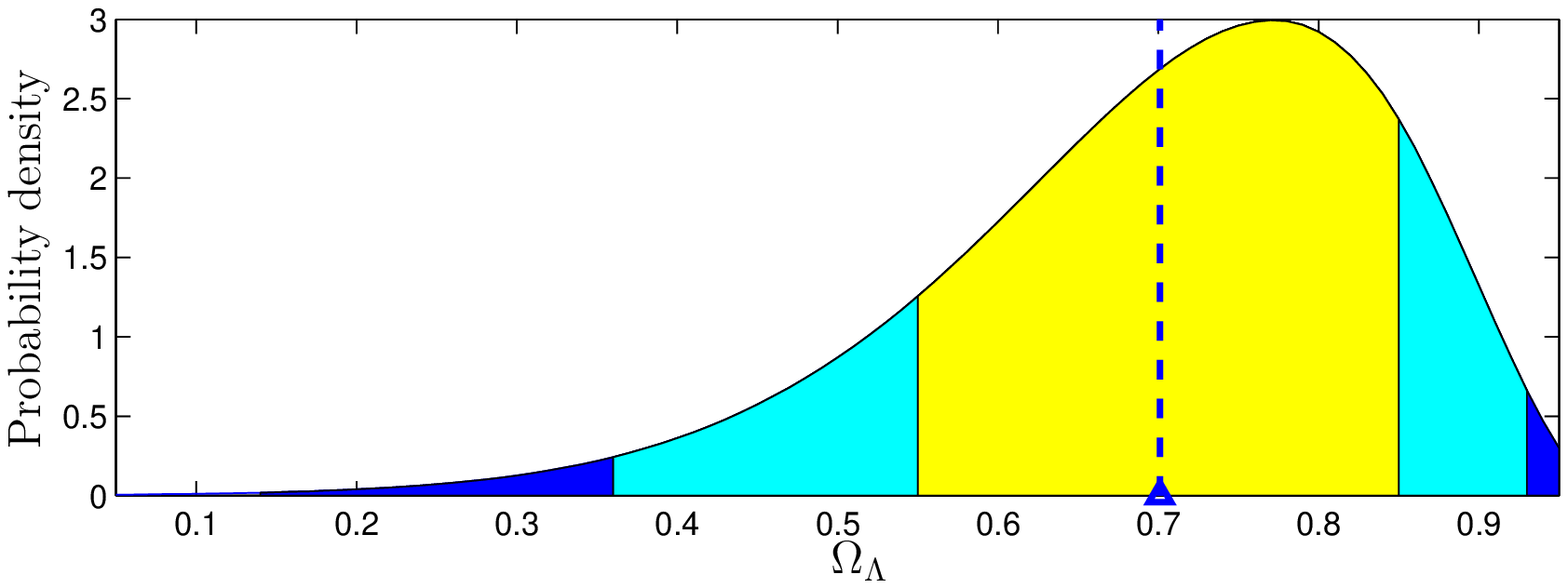}}
\subfigure[Likelihood distribution for \olambda\ obtained using \sbw]{\includegraphics[width=\pdfwidth]{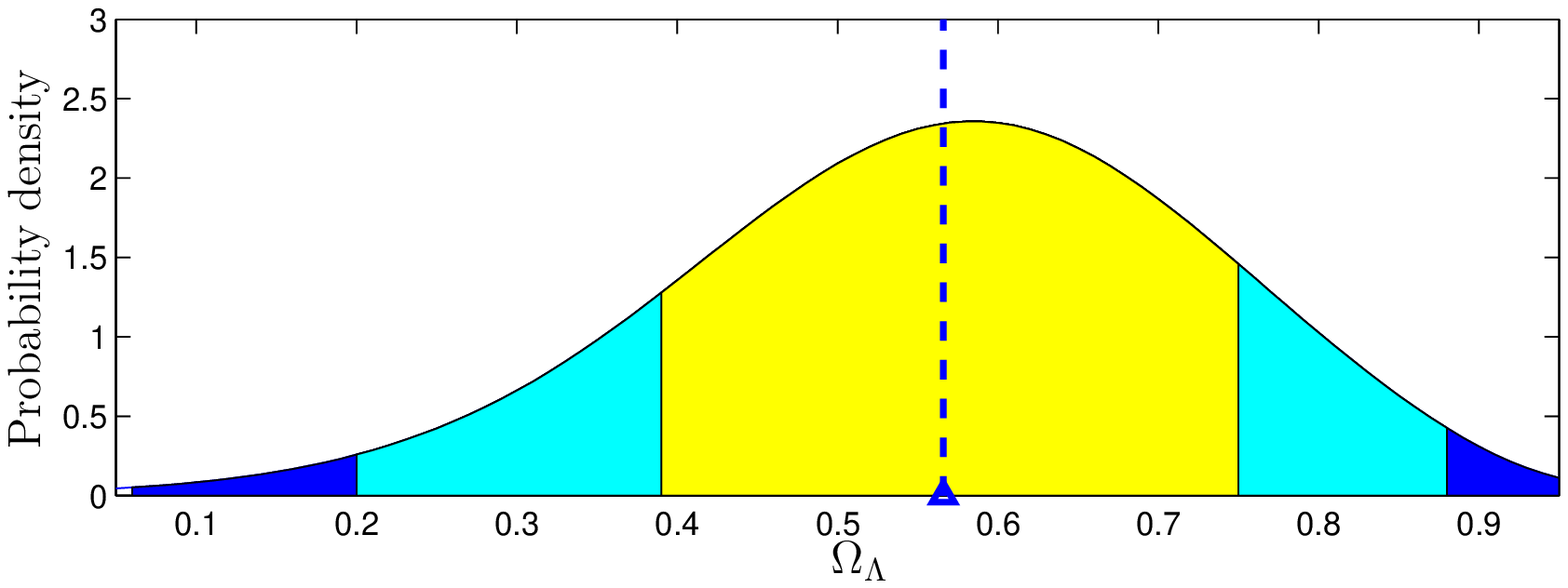}}
\subfigure[Cumulative probability functions for $P(\olambda>x)$ ]{\includegraphics[width=\pdfwidth]{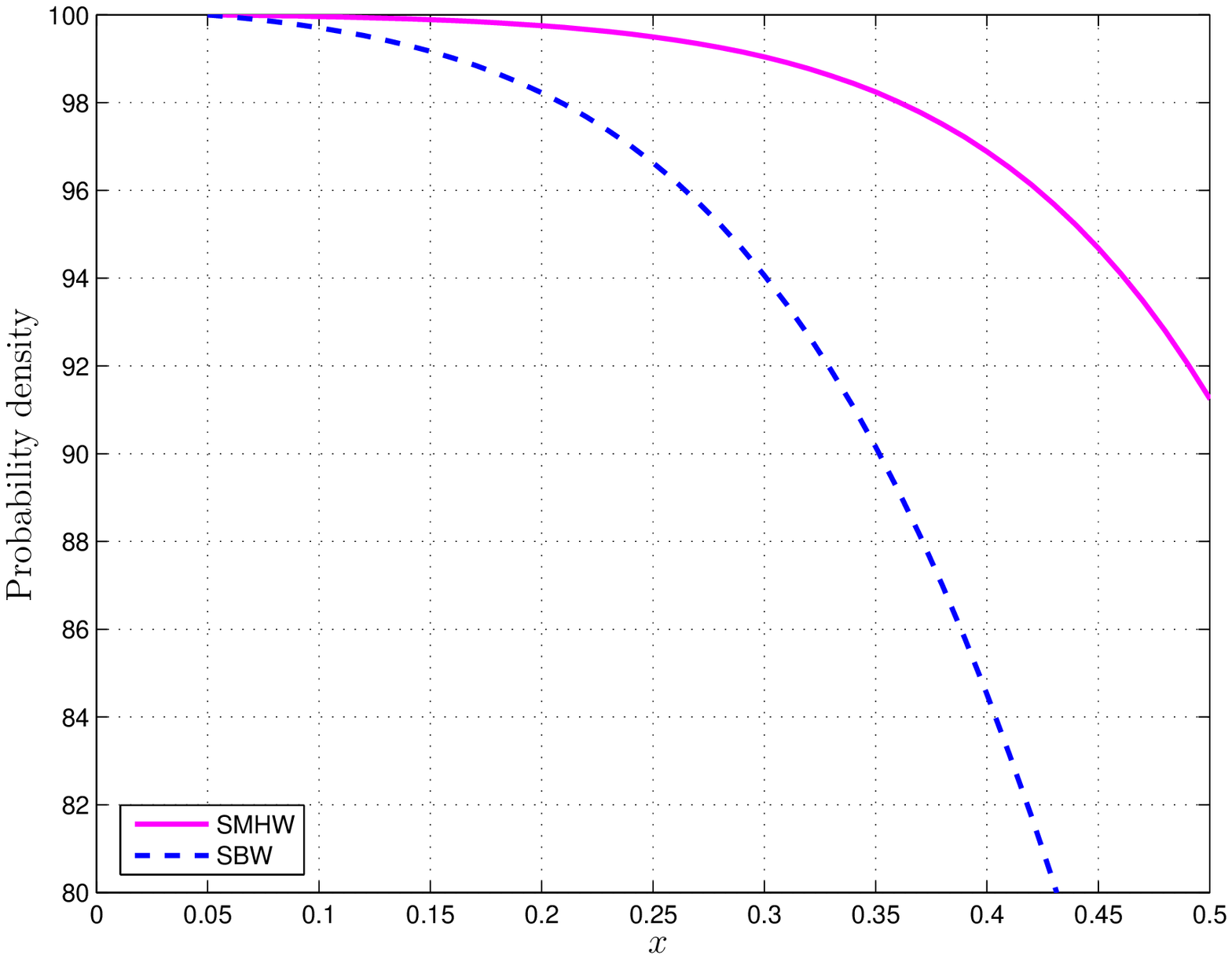}}
\caption{Likelihood distributions and cumulative probability functions for \olambda\ when $\w=-1$.
Confidence regions at 68\% (yellow/light-grey), 95\% (light-blue/grey) and 99\% (dark-blue/dark-grey) are also shown on the likelihood distributions, with the parameter estimates made from the mean of the distribution shown by the triangle and dashed line.  The cummulative probability functions show the probability $P(\olambda>x)$ for the \smhw\ (solid magenta curve) and \sbw\ (dashed blue curve).
}
\label{fig:pdfwone}
\end{figure}

It is also interesting to use our detection of the \isw\ effect to provide an estimate of the overall evidence for the existence of dark energy when assuming a pure cosmological constant.  We compute the likelihood distribution of just the vacuum energy density \olambda\ in the presence of a cosmological constant by repeating the procedure described in \sectn{\ref{sectn:constraintsboth}} for $\baypar=\olambda$ and $\w=-1$.  This essentially reduces to taking a slice through the 2-dimensional likelihood surfaces shown in \fig{\ref{fig:pdfall}} at $\w=-1$.  The corresponding likelihood distributions for each wavelet are shown in \fig{\ref{fig:pdfwone}}.  
From the mean of the distributions we obtain the following vacuum energy density estimates:
$\olambda=0.70_{-0.15}^{+0.15}$ using the \smhw ;
$\olambda=0.57_{-0.18}^{+0.18}$ using the \sbw.  
Within error bounds these parameter estimates are again consistent with each other and with estimates made using other techniques and data sets.

We show also in \fig{\ref{fig:pdfwone}} the cumulative probability \mbox{$P(\olambda>x)$}.  One may read directly  off this plot the evidence for the existence of dark energy above a certain level.  For instance, using the \smhw\ we may state that $\olambda>0.1$ at the $99.96\%$ level.  Using the \sbw\ we may state that $\olambda>0.1$ at the $99.71\%$ level.  The cummulative probability function falls off faster for the \sbw , indicating that the \smhw\ happens to be more effective at ruling out a low vacuum energy density (\ie\ the \smhw\ predicts a higher \olambda).
In any case, we have very strong evidence for dark energy in the case of a pure cosmological constant.

\section{Conclusions}
\label{sec:conclusions}

We have performed a directional spherical wavelet analysis to search for correlations between the \wmap\ and \nvss\ data arising from the \isw\ effect.
Wavelets are an ideal tool for searching for the \isw\ effect due to the localised nature of the effect and the localisation afforded by a wavelet analysis.  Furthermore, we use directional spherical wavelets that allow one to probe not only scale and spatially localised structure but also orientated structure in the data.

Using this technique we have detected a significant correlation between the \wmap\ and \nvss\ data.  The detection is made using both the directional \smhw\ and the \sbw , on approximately the same wavelet scale of 
$(\scalea,\scaleb)=(100\arcmin,300\arcmin)$, corresponding to an overall effective size on the sky of approximately $10\degrees$.  The scale and positive sign of the detected correlation is consistent with the \isw\ effect.  To test this hypothesis further we examined other possible causes of the signal and after performing some preliminary tests ruled out both foreground contamination and systematics in the \wmap\ data.  A wavelet analysis inherently provides localised information, thus we were able to localise those regions on the sky that contribute most strongly to the overall correlation.  On removing these regions from the analysis the correlation originally detected is reduced in significance, as one would expect, but it is not eliminated, remaining at the 95\% significance level in some cases.  It therefore appears that although the localised regions that we detected are a strong source of correlation, they are not the sole source.  This is again consistent with predictions made using the \isw\ effect, where one would expect to observe weak correlations over the entire sky rather than just a few localised regions.  Finally, we examined 18 of the most significant localised regions in more detail in high-resolution maps of the \nvss\ and Bonn 1420MHz radio surveys.  These regions are not atypical and do not, in general, correspond to regions with particularly bright point sources that could have caused contamination of the data.  All of these findings suggest that the correlation that we have detected is due to the \isw\ effect.
In our (nearly) flat universe the \isw\ effect is present only in the presence of dark energy, hence our detection of the effect is direct evidence for dark energy.  

We have used our detection of the \isw\ effect to constrain dark energy parameters by comparing theoretical predictions made by different cosmological models with the data.
Constraining both the vacuum energy density \olambda\ and the equation-of-state parameter \w, we obtained the following parameter estimates:
$\olambda=0.63_{-0.17}^{+0.18}$, $\w=-0.77_{-0.36}^{+0.35}$ using the \smhw ;
$\olambda=0.52_{-0.20}^{+0.20}$, $\w=-0.73_{-0.46}^{+0.42}$ using the \sbw.
We have therefore obtained independent estimates of the parameters that, within error bounds, are consistent with one another and also with estimates made from many other analysis procedures and data sets.
The errors we obtain on the parameter estimates are relatively large.  Although wavelets perform very well when attempting to detect the \isw\ effect since one may probe particular scales, positions and orientations, once all information is incorporated when computing the parameter estimates the performance of the wavelet analysis is comparable to other linear techniques, as expected.
We also constrain \olambda\ for the case of a pure cosmological constant, \ie\ $\w=-1$.  For this case we obtain the following parameter estimates:
$\olambda=0.70_{-0.15}^{+0.15}$ using the \smhw ;
$\olambda=0.57_{-0.18}^{+0.18}$ using the \sbw.  
Again the estimates are consistent with one another and with those made elsewhere.
For this case we also examined the evidence for a non-zero cosmological constant, ruling out $\olambda < 0.1$ at greater than the 99\% level using each wavelet.

In this work we have extended to directional wavelets the spherical wavelet analysis technique used to probe \cmb\ data and \lss\ tracers for cross-correlations.  The effectiveness of the wavelet analysis technique is illustrated herein by the highly significant detection of the \isw\ effect that we are able to make using the \wmap\ and \nvss\ data.  In addition, the spherical wavelet covariance approach to detecting correlations can be applied to other applications, such as searching for the Sunyaev-Zel'dovich effect \citep{hansen:2005} or looking for common structures between the \cmb\ and foregrounds \citep{liu:2006}.
It would also be interesting to repeat our analysis using different tracers of the \lss\ and also the next release of the \wmap\ data to see if the \isw\ signal that we have detected remains.


\section*{Acknowledgments}

We thank Jacques Delabrouille and Daniel Mortlock for insightful discussions and comments on the manuscript and Dave Green for help accessing the \nvss\ and Bonn 1420MHz radio survey catalogues.  JDM would like to thank all at Universidad de Cantabria and Coll\`{e}ge de France for the warm hospitality he received throughout his visit to each.  JDM also thanks the Association of Commonwealth Universities and the Cambridge Commonwealth Trust for the support of a Commonwealth (Cambridge) Scholarship, King's College Cambridge for a Ferris Travel Grant and the Cambridge Philosophical Society for a travel grant.  PV and EMG acknowledge financial support from the Spanish MEC project \mbox{ESP2004-07067-C03-01}.  Some of the results in this paper have been derived using the \healpix\ package \citep{gorski:2005}.  We acknowledge the use of the \lambdaarchtext\ (\lambdaarch).  Support for \lambdaarch\ is provided by the NASA Office of Space Science.


%
%



\appendix
\section{Theoretical directional spherical wavelet covariance}
\label{appn:theo_covar}

When using azimuthally symmetric wavelets the spherical harmonic coefficients of the spherical wavelet transform may simply be written as the product of the signal and wavelet harmonic coefficients.  This is not the case when using directional wavelets (\ie\ wavelets that are not azimuthally symmetric).  For this reason the theoretical form of the azimuthally symmetric wavelet covariance estimator derived by \citet{vielva:2005} is not trivially extended to directional wavelets.  We derive here the theoretical directional spherical wavelet covariance.  For simplicity, and without loss of generality, we ignore beam and pixel window functions in the analysis (which, in any case, may easily be put in by hand at the end if required).

Firstly, we derive an expression for the real space angular correlation function:
\begin{eqnarray}
< \nd(\sa) \: \tp^\cconj(\sa\p) > &=&
\sum_{\el=0}^\infty \clnttheo \: \sum_{\m=-\el}^\el
\shfarg{\el}{\m}{\sa} \:
\shfargc{\el}{\m}{\sa\p} \nonumber \\
&=&
\sum_{\el=0}^\infty
\frac{2\el+1}{4\pi} \:
\leg{\el}{\sa \dotprod \sa\p} \:
\clnttheo
\spcend ,
\label{eqn:ccf}
\end{eqnarray}
where we have made use of \eqn{\ref{eqn:cltheo}} and also the addition theorem for spherical harmonics:
\begin{equation}
\sum_{\m=-\el}^\el
\shfarg{\el}{\m}{\sa} \:
\shfargc{\el}{\m}{\sa\p} =
\frac{2\el+1}{4\pi} \:
\leg{\el}{\sa \dotprod \sa\p}
\spcend ,
\end{equation}
where $\shfarg{\el}{\m}{\sa}$ are the spherical harmonic functions and $\leg{\el}{x}$ are the Legendre polynomials.

Using \eqn{\ref{eqn:ccf}} we may write the theoretical wavelet covariance as
\begin{eqnarray}
\lefteqn{< \skywav_\wav^\ndlab(\scaleab,\eul) \:
\skywav_\wav^{\tplab\cconj}(\scaleab,\eul) >} \nonumber \\
&=&
\int_{\sphere}
\int_{\sphere}
\dmu{\sa}
\dmu{\sa\p} \:
\wav_{\scaleab,\eul}^\cconj(\sa) \:
\wav_{\scaleab,\eul}(\sa\p) \: \nonumber \\
&& \:\:\: \times \:
< \nd(\sa) \: \tp^\cconj(\sa\p) > \nonumber \\
&=&
\sum_{\el=0}^\infty \frac{2\el+1}{4\pi} \:
\clnttheo
\int_{\sphere}
\int_{\sphere}
\dmu{\sa} 
\dmu{\sa\p} \: \nonumber \\
&& \:\:\: \times \:
\wav_{\scaleab,\eul}^\cconj(\sa) \:
\wav_{\scaleab,\eul}(\sa\p) \:
\leg{\el}{\sa \dotprod \sa\p}
\spcend .
\label{eqn:appn_cov1}
\end{eqnarray}
Assuming ergodicity, we may relate \eqn{\ref{eqn:appn_cov1}} directly to our wavelet covariance estimator.  Furthermore, under our assumption that both the \cmb\ and galaxy density fields are isotropic, the wavelet covariance is independent of our choice of the position and orientation of the wavelet, hence we may set the rotation of the wavelet to zero, \ie\ $\eul=\mathbf{0}$.  Thus we may write
\begin{equation}
\wcov^{\ndlab\tplab}(\scaleab, \eulc) =
\sum_{\el=0}^\infty
\frac{2\el+1}{4\pi} \:
\clnttheo
\int_{\sphere}
\dmu{\sa} \:
\covsubterm(\sa) \:
\wav_{\scaleab}^\cconj(\sa)
\label{eqn:appn_cov2}
\end{equation}
where
\begin{equation}
\covsubterm(\sa) =
\int_{\sphere}
\dmu{\sa\p} \:
\leg{\el}{\sa \dotprod \sa\p} \:
\wav_{\scaleab}(\sa\p)
\spcend .
\end{equation}
Notice that the theoretical wavelet covariance specified by \eqn{\ref{eqn:appn_cov2}} is independent of the orientation of the analysis \eulc.  The assumption of isotropy does not imply that the fields considered cannot contain localised anisotropic structure, thus individual wavelet covariance estimators obtained from the data will vary over orientations, but the theoretical prediction for each orientation will be the same when we consider the statistics of the aggregate fields.  Essentially, we get a number of samples of a statistic for which we have a single theoretical prediction, thereby actually increasing the performance of any subsequent comparison between the data and theoretical predictions.  Directional wavelets are therefore still advantageous in probing localised oriented structure in the data.

Now we consider the spherical harmonic coefficients of $\covsubterm(\sa)$.  Since $\leg{\el}{\sa \dotprod \sa\p}$ is azimuthally symmetric it should be possible to write the spherical harmonic coefficients of $\covsubterm(\sa)$ simply in terms of the harmonic coefficients of the wavelet:
\begin{eqnarray}
\covsubterm_{\el\p\m\p}
&=&
\int_{\sphere}
\dmu{\sa} \:
\covsubterm(\sa) \:
\shfargc{\el\p}{\m\p}{\sa} \nonumber \\
&=&
\sum_{\el\pp=0}^\infty \:
\sum_{\m\pp=-\el\pp}^{\el\pp}
(\wav_{\scaleab})_{\el\pp\m\pp} \:
I_{\el\pp\el\p\m\pp\m\p}^\el
\spcend ,
\end{eqnarray}
where
\begin{eqnarray}
I_{\el\pp\el\p\m\pp\m\p}^\el &=&
\int_{\sphere}
\int_{\sphere}
\dmu{\sa} \:
\dmu{\sa\p} \:
\leg{\el}{\sa \dotprod \sa\p} \:
\shfargc{\el\p}{\m\p}{\sa} \:
\shfarg{\el\pp}{\m\pp}{\sa\p}
\nonumber \\
&=&
\int_{\sphere}
\int_{\sphere}
\dmu{\sa} \:
\dmu{\sa\p} \:
\frac{4\pi}{2\el+1} \nonumber \\
&& \:\:\: \times \:
\sum_{\m=-\el}^\el
\shfarg{\el}{\m}{\sa} \:
\shfargc{\el}{\m}{\sa\p} \:
\shfargc{\el\p}{\m\p}{\sa} \:
\shfarg{\el\pp}{\m\pp}{\sa\p}
\nonumber \\
&=&
\frac{4\pi}{2\el+1} \:
\kron_{\el\el\p}
\kron_{\el\el\pp}
\kron_{\m\p\m\pp}
\spcend ,
\end{eqnarray}
where we make use of the addition theorem for spherical harmonics again.  The last line follows from the orthogonality of the spherical harmonics:
\begin{equation}
\int_{\sphere}
\dmu{\sa} \:
\shfarg{\el}{\m}{\sa} \:
\shfarg{\el\p}{\m\p}{\sa}
= \kron_{\el\el\p} \kron_{\m\m\p}
\spcend .
\label{eqn:shorthog}
\end{equation}
Thus, the harmonic coefficients of $\covsubterm(\sa)$ are given by
\begin{equation}
\covsubterm_{\el\p\m\p} =
\frac{4\pi}{2\el+1} \:
(\wav_{\scaleab})_{\el\p\m\p} \:
\kron_{\el\el\p}
\spcend .
\end{equation}

We are now in a position to derive the final expression for the theoretical wavelet covariance in terms of the wavelet spherical harmonic coefficients and the theoretical \cmb-galaxy density cross-power spectrum.  Expanding $\covsubterm(\sa)$ in \eqn{\ref{eqn:appn_cov2}} into its spherical harmonic decomposition yields
\begin{eqnarray}
\wcov^{\ndlab\tplab}(\scaleab, \eulc)
&=&
\sum_{\el=0}^\infty
\clnttheo \:
\sum_{\m=-\el}^\el \:
(\wav_{\scaleab})_{\el\m} 
\int_{\sphere}
\dmu{\sa} \:
\shfarg{\el}{\m}{\sa} \:
\wav_{\scaleab}^\cconj(\sa) \nonumber \\
&=&
\sum_{\el=0}^\infty
\clnttheo \:
\sum_{\m=-\el}^\el \:
\left| (\wav_{\scaleab})_{\el\m}  \right|^2
\spcend ,
\label{eqn:appn_cov3}
\end{eqnarray}
where the last line follows from representing $\wav_{\scaleab}^\cconj(\sa)$ by its harmonic expansion and noting again the orthogonality of the spherical harmonics \eqn{\ref{eqn:shorthog}}.
For the case of azimuthally symmetric wavelets the \m-modes of the spherical harmonic coefficients of the wavelet are non-zero only for $\m=0$, thus \eqn{\ref{eqn:appn_cov3}} reduces to the form given by \citet{vielva:2005} (the $({2\el+1})/{4\pi}$ discrepancy arises since \citet{vielva:2005} use Legendre, rather than spherical harmonic coefficients).
We have therefore obtained a simple form, that may be computed easily, for the theoretical wavelet covariance of directional wavelets in terms of the wavelet spherical harmonic coefficients and the cross-power spectrum.


\label{lastpage}

\end{document}